# Different dielectric, magnetic, and magnetodielectric mechanisms in *M*-type $BaFe_{12}O_{19}$ hexaferrite regulated by doping $Ga^{3+}$ and $In^{3+}$ cations

Yang Yang, Run-Yu Lei, Jian-Ping Zhou,* and Xiao-Ming Chen

*School of Physics and Information Technology, Shaanxi Normal University, Xi'an 710119, People's Republic of China*

**Abstract**: *M*-type hexaferrite is a material with rich physical contents, such as magnetism, dielectric property, and magnetodielectric (MD) effect. In this article, we systematically investigated the magnetic, dielectric, and MD properties of $BaFe_{12-x}Me_xO_{19}$ ($Me$ = Ga and In; $x$ = 0.0, 1.2, 1.8, and 2.4) ceramics prepared by a solid-state reaction method. The $Ga^{3+}$ cations with a smaller radius preferentially substitute the $Fe^{3+}$ ions in $FeO_6$ octahedra while the $In^{3+}$ cations with a larger radius tend to replace the $Fe^{3+}$ ions in $FeO_5$ bipyramids of *R* blocks, inducing different physical characteristics. The pure $BaFe_{12}O_{19}$ and Ga-doped samples show ferrimagnetism in the temperature range from 10 K to 300 K. The In-doped samples exhibit a transition from non-collinear magnetism to collinear ferrimagnetism at 39 K, 128 K, and 144 K for the doping amount of $x$ = 1.2, 1.8, and 2.4, respectively. The dielectric decrease of pure $BaFe_{12}O_{19}$ at around 10–175 K is attributed to the quantum paraelectric state, and the shoulder peaks of tan $\delta$ at about 140–200 K are from electron hopping. The dipole glass state is responsible for the dielectric peak of Ga-doped samples at around 20–40 K. The dielectric increase and plateau of In-doped samples are mainly ascribed to the electron hopping at low temperatures. Their dielectric properties at high temperatures are all attributed to the interfacial polarization caused by the Maxwell-Wagner effect. The MD effect also has different origins for the various samples at low temperatures. For the pure $BaFe_{12}O_{19}$, the negative MD effect at extremely low temperatures and the positive MD effect after warming are ascribed to spin-phonon coupling and field-dependent electron hopping, respectively. The positive MD effect in Ga-doped hexaferrites results from the field-dependent electric dipoles inside $FeO_5$ bipyramids. For the In-doped samples, the negative MD effect and subsequent transformation to the positive MD effect originate from the field-dependent non-collinear spin ordering and electron hopping, respectively. The MD effect at high temperatures is attributed to the combination of magnetoresistance and Maxwell-Wagner effects. These research results are helpful for understanding the relationship among doped ions, spin order, dielectric property, and MD effect in the *M*-type hexaferrite.

* Corresponding author. E-mail address: zhoujp@snnu.edu.cn, ORCID: 0000-0003-0807-1404, ResearcherID: AGE-2972-2022.

## I. Introduction

The magnetodielectric (MD) effect, defined as the modulation of dielectric behavior by an external magnetic field, is a significant research topic in multiferroics [1,2]. There are two types of multiferroics. The first group is called type-I multiferroic, in which ferroelectricity and magnetism have different sources. The polarization in type-I multiferroic mainly originates from $d^0$ ferroelectricity, $6s^2$ lone pair, and geometric frustration [3]. Another is type-II multiferroic, in which the particular spin order causes polarization, implying a strong coupling between ferroelectricity and magnetism [4]. In general, the polarization in type-II multiferroic derives from three main mechanisms: inverse Dzyaloshinskii-Moriya interaction of non-collinear spin order, exchange striction of "up-up-down-down" spin order, and *p*-*d* orbital hybridization [2,5]. There are four main MD origins in multiferroics. The magnetoelectric (ME) effect can induce the MD effect as an intrinsic source in single multiferroics. The ME effect can be achieved through the interaction between spin order and spontaneous polarization. The direction or intensity of polarization will be modulated by an external field through the spin-lattice or spin-orbit coupling, leading to an ME effect. Meanwhile, the associated dielectric permittivity will change with the evolution of polarization, resulting in the MD effect [6]. In addition, the MD effect can be achieved by the magnetic field-dependent electron hopping [7], and the spin-phonon coupling [8]. The magnetoresistance effect combined with the Maxwell-Wagner effect is another way to inspire the MD effect. This source is considered an extrinsic mechanism because it is unrelated to the ME effect [9].

Hexaferrite enjoys high research value in the ME and MD effects due to the strong coupling between magnetic order and electric dipoles in a wide temperature range [10-14]. There are six types of hexaferrites, i.e., *M*, *Y*, *Z*, *X*, *U*, and *W* types, according to their crystal structures [15]. The *M*-type hexaferrite ($AFe_{12}O_{19}$: $A$ = Ba, Sr, Pb, Ca, etc.) has a relatively simple crystal structure and unique physical features among them. Its unit cell is constituted by different kinds of oxygen polyhedrons to locate $Fe^{3+}$ cations as shown in Fig. 1(a). There are 5 sites marked with different colors, i.e., 12*k* site inside $FeO_6$ octahedron, 2*a* site inside $FeO_6$ octahedron, $4f_2$ site inside $FeO_6$ octahedron, 2*b* site inside $FeO_5$ bipyramid, and $4f_1$ site inside $FeO_4$ tetrahedron. The basic crystal units are *S* block ($2MeFe_2O_4$, *Me* = divalent metal ion) and *R* block ($BaFe_6O_{11}$), and similar *S** and *R** blocks, which are formed by rotating *S* and *R* blocks by 180° around the *c*-axis [16].

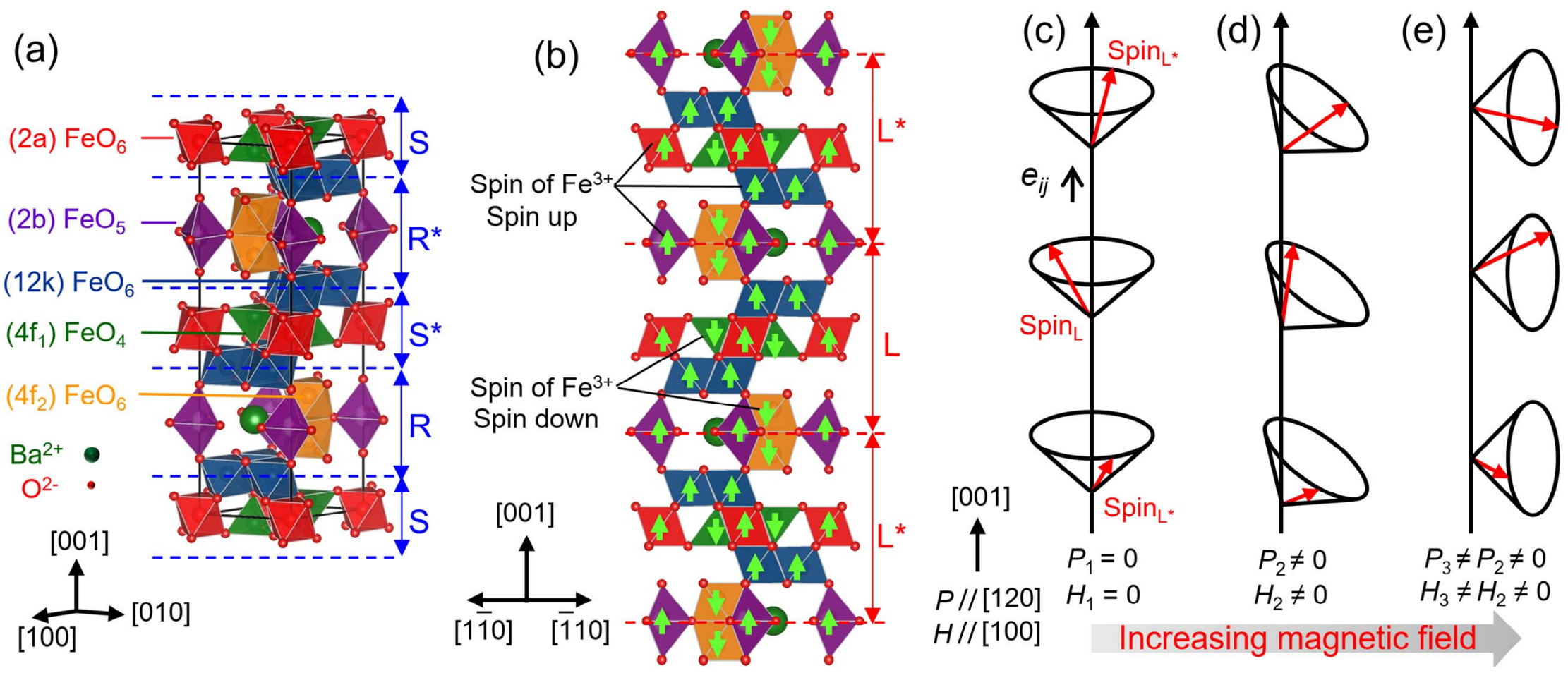

FIG. 1. (a) Schematic of $BaFe_{12}O_{19}$ crystal structure, (b) magnetic structure of cross-section view in (1 1 0) plane, and (c)–(e) schematics of non-collinear spin order for the *M*-type hexaferrite under different magnetic fields. $\boldsymbol{e_{ij}}$ is the unit vector connecting the adjacent basic magnetic units.

The pure *M*-type hexaferrite shows ferrimagnetism because there are more $Fe^{3+}$ cations in spin-up sites [$12Fe^{3+}$ (12*k*) + $2Fe^{3+}$ (2*a*) + $2Fe^{3+}$ (2*b*)] than those in spin-down sites [$4Fe^{3+}$ ($4f_1$) + $4Fe^{3+}$ ($4f_2$)] along the *c*-axis of unit cell [17,18] as shown in Fig. 1(b). The pure $BaFe_{12}O_{19}$ has a calculated net magnetic moment of 40 μB per unit cell, i.e., 20 μB per formula [15] due to the magnetic moment of 5 μB for $Fe^{3+}$ cation. Various multiferroic effects can be achieved due to its unique crystal and magnetic structures. One important discovery is the non-collinear spin ordering-mediated ME and MD effects [10,19-21]. The block-type magnetic structure can be established by changing the original magnetic interaction or uniaxial magnetocrystalline anisotropy through element doping. Then, the net magnetic direction of the adjacent basic magnetic unit *L* and *L** blocks will deviate from the c-axis, forming a non-collinear longitudinal conical spin order as shown in Figs. 1(b) and 1(c). The conical axis of spin cone will tilt under an external magnetic field and further change polarization as shown in Figs. 1(d) and 1(e) due to the inverse Dzyaloshinskii-Moriya interaction, producing a spin ordering-mediated ME effect. Consequently, the spin ordering-mediated MD effect can be achieved on the basis of the ME effect [20-23].

Other significant research works found the MD effect could exist in the *M*-type hexaferrite, but in which the spin order is independent of electric dipoles, maybe belonging to the type-I multiferroic. There are three possible origins for this type of MD effect and electric dipole. The electric dipole exists in $FeO_5$ bipyramid of pure $BaFe_{12}O_{19}$. The inner $Fe^{3+}$ cations in bipyramids will occupy the off-center positions along the *c*-axis (two 4*e* positions) with a lower energy, rather than the center of bipyramid (2*b* position) with a higher energy, causing an electric dipole. The strong quantum fluctuation in pure $BaFe_{12}O_{19}$ induces a dynamic occupation between two 4*e* positions for the $Fe^{3+}$ cations inside bipyramids, resulting in the centrosymmetric space group of $BaFe_{12}O_{19}$ [24,25]. Thus, $BaFe_{12}O_{19}$ exhibits a typical quantum paraelectricity [26]. Another different result revealed the inner $Fe^{3+}$ cation moves to an off-center position and two apical $O^{2-}$ anions leave the original positions in a certain octahedron. The $Fe^{3+}$ cation shifts along the *b*-axis, while the $O^{2-}$ anions shift in the opposite direction of the *a*-axis, forming an electric dipole in this distorted octahedron and leading to the non-centrosymmetric space group of $BaFe_{12}O_{19}$ [27]. But the doped *M*- type hexaferrite is well described within the non-centrosymmetric space group on the basis of the refinement of neutron diffraction patterns. The local crystallographic distortion occurs in a tetrahedron, bipyramid, and all three kinds of octahedra, together contributing to an electric dipole of these doped hexaferrites [28]. These electric dipoles or polarization provide the foundation of MD effect in *M*-type hexaferrite, and element doping is an effective method to adjust the crystal structure and MD effect.

Beyond the above two parts, the various chemical states of iron in *M*-type hexaferrite will give rise to electron hopping and Maxwell-Wagner polarization, which causes the relevant MD effect [23,29]. In addition, the spin-phonon coupling in *M*-type hexaferrite changes the dielectric permittivity under a magnetic field [30,31]. Therefore, the *M*-type hexaferrite is an attractive system in MD research. The lattice distortion from element doping will induce the non-collinear conical spin order, further producing the MD effect [32-34]. The $Ga^{3+}$ cation with a smaller radius and $In^{3+}$ cation with a larger radius are appropriate to regulate the crystal and magnetic structures of the *M*-type hexaferrite for comparative investigations. In this article, we investigated the origin of MD effect in Ga- and In-doped *M*-type $BaFe_{12}O_{19}$ hexaferrites through systematic research on their crystal structure, magnetic, dielectric, and MD properties. These results provide a foundamental understanding of the MD effect in *M*-type hexaferrite.

## II. Experimental details

### A. Sample preparation

The polycrystalline samples $BaFe_{12-x}Me_xO_{19}$ ($Me$ = Ga and In; $x$ = 0.0, 1.2, 1.8, and 2.4) were prepared by the solid-state reaction method. Stoichiometric raw powders of $BaCO_3$, $Fe_2O_3$, and $In_2O_3$/$Ga_2O_3$ with a purity of 99.9% were thoroughly mixed. The mixed powders were calcined at 1125 °C for 6 hours in the air. The calcined powders were shaped into pellets by static compaction. The pellets were sintered at 1230 °C for 60 hours in an oxygen atmosphere to obtain the ceramics and annealed at 1000 °C for 50 hours in an oxygen atmosphere to reduce the oxygen defects.

### B. Measurements

The X-ray diffraction (XRD) was tested by a diffractometer system (Rigaku, D/Max 2550). The Raman spectra were determined by a spectrometer system (Horiba Jobin Yvon, LabRAM HR Evolution). The X-ray photoelectron spectroscopy (XPS) was obtained by a spectrometer system (Kratos Analytical, AXIS ULTRA). The thermomagnetic curves and hysteresis loops were collected on a vibrating sample magnetometer (Cryogenic, CFMS-8T) to investigate the magnetic properties. The dielectric permittivity under different temperatures and different magnetic fields was measured by an MD testing system. In this system, the dielectric permittivity was collected by an impedance analyzer (Keysight, E4980AL). The temperatures from 10 K to 300 K were controlled by a temperature controller (East Changing Technologies, TC280). The magnetic fields from −50 kOe to 50 kOe were provided by a superconducting magnet (Cryogenic, CFM-6T-150-RT). The temperature dependence of dielectric permittivity at different frequencies was also tested by this MD system. The dielectric permittivity, dielectric loss (tan $\delta$), and imaginary of the complex electric modulus ($M''$) in a broad frequency range of 1 Hz–100 kHz from 120 K to 300 K were obtained by a broadband dielectric spectrometer (Novocontrol, Alpha-A) to further investigate the dielectric property. The impedance spectroscopy at room temperature was measured by an impedance analyzer (Agilent, 4294A).

## III. Results and discussion

### A. Crystal structure and chemical analysis

Fig. 2(a) shows the XRD patterns of $BaFe_{12-x}Me_xO_{19}$ ($Me$ = Ga and In; $x$ = 0, 1.2, 1.8, and 2.4) ceramics at room temperature. The XRD peaks well match the $BaFe_{12}O_{19}$ phase (PDF No: 84-0757, space group: $P6_3/mmc$). The main peaks shift to a lower angle after doping $In^{3+}$ cations with a larger radius (0.80 Å) and to a little higher angle after doping $Ga^{3+}$ cations with a smaller radius (0.63 Å) in comparison with the $Fe^{3+}$ cation radius (0.64 Å). Correspondingly, the hexaferrite unit cell shrinks after doping $Ga^{3+}$ cations but expands after doping $In^{3+}$ cations. Both lattice parameters $a$ and $c$ increase with the doping amount for the In-doped samples, but *a* reduces and *c* increases for the Ga-doped samples as shown in Fig. 2(b) and Table I.

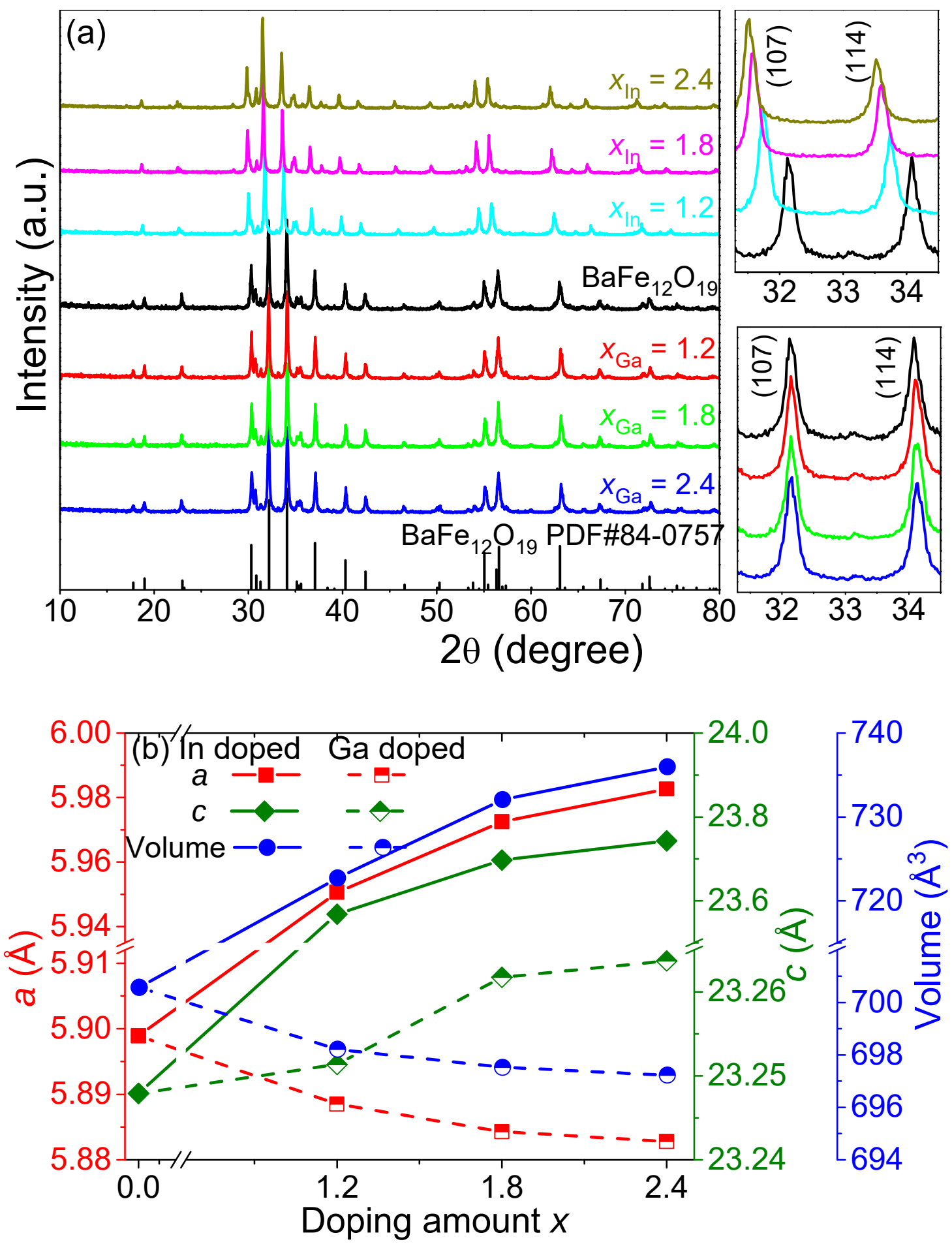

FIG. 2. (a) XRD patterns of $BaFe_{12-x}Me_xO_{19}$ (*Me* = Ga and In; *x* = 0, 1.2, 1.8, and 2.4) ceramics, and their subtle patterns around the main peaks. (b) Dependence of lattice parameters and cell volume on the doping amount.

TABLE I. Summary of the physical quantities of $BaFe_{12-x}Me_xO_{19}$ ($Me$ = Ga and In; $x$ = 0, 1.2, 1.8, and 2.4)

| Sample | $x$ | Crystal structure | | | Raman spectrum | Magnetic property | Dielectric property | | Magnetodielectric (MD) effect | | | |
|---|---|---|---|---|---|---|---|---|---|---|---|---|
| | | $a$ (Å) | $c$ (Å) | $v$ (Å$^3$) | | | Low temperature | Close to room temperature | Low temperature | Close to room temperature | Max coefficient | Max coefficient at zero field |
| Pure | – | 5.90 | 23.25 | 700.58 | – | Ferrimagnetism | Quantum paraelectric, and electron hopping | Maxwell-Wagner effect | Negative MD effect from spin-phonon coupling, positive MD effect from the field-dependent electron hopping | Extrinsic oscillating MD behavior | – | – |
| Ga-doped | 1.2<br>1.8<br>2.4 | 5.89<br>5.88<br>5.88 | 23.25<br>23.26<br>23.26 | 698.21<br>697.53<br>697.23 | $Ga^{3+}$ replaces $Fe^{3+}$ in $FeO_6$ octahedra of $R$ blocks | Ferrimagnetism | Electric dipole glass state | Maxwell-Wagner effect | Positive MD effect from the field-dependent electric dipoles inside $FeO_5$ bipyramids | Extrinsic oscillating MD behavior | 0.147 (at 50 K)<br>0.215 (at 50 K)<br>0.252 (at 50 K) | 0.146 (at 50 K)<br>0.215 (at 50 K)<br>0.252 (at 50 K) |
| In-doped | 1.2<br>1.8<br>2.4 | 5.95<br>5.97<br>5.98 | 23.57<br>23.70<br>23.74 | 722.74<br>732.06<br>735.95 | $In^{3+}$ replaces $Fe^{3+}$ in $FeO_5$ bipyramids of $R$ blocks | Noncollinear magnetism (10 K–$T_{M2}$),ferrimagnetism ($T_{M2}$–300 K) | Electron hopping | Maxwell-Wagner effect | Negative MD effect from the field-dependent noncollinear spin ordering, positive MD effect from the field-dependent electron hopping | Extrinsic oscillating MD behavior | 0.214 (at 50 K)<br>0.432 (at 100 K)<br>0.501 (at 100 K) | 0.213 (at 50 K)<br>0.431 (at 100 K)<br>0.502 (at 100 K) |

Fig. 3 shows the Raman spectra of $BaFe_{12-x}Me_xO_{19}$ ceramics at room temperature. There are 7 notable peaks for the pure $BaFe_{12}O_{19}$ sample. The peaks at around 337.9, 412.2, 468.7, and 524.8 $cm^{-1}$ are marked as I, II, III, and IV, respectively. The second strongest peak, strongest peak, and shoulder peak at around 617.6, 687.9, and 726.9 $cm^{-1}$ are marked as V, VI, and VII, respectively. The Fe-O bonds in different oxygen polyhedrons produce 42 Raman active modes ($11A_{1g}$, $14E_{1g}$, and $17E_{2g}$). Peaks V and VI are related to the Fe-O bonds of $FeO_6$ octahedra and $FeO_5$ bipyramids in the *R* blocks, respectively [35]. Peak VII is associated with the Fe-O bonds of $FeO_4$ tetrahedrons in the *S* blocks. The peak II is correlated with the Fe-O bonds of $FeO_6$ octahedra at the boundary of *R* and *S* blocks. Peak III is assigned to the Fe-O bonds of $FeO_6$ octahedra at the boundary of two blocks (*R* and *S*) and $FeO_6$ octahedra in the *S* blocks. Peak I is attributed to the Fe-O bonds in all $FeO_6$ octahedra. In addition, peak IV arises from the $E_{1g}$ active mode, while other peaks originate from the $A_{1g}$ active mode. The Raman spectra are clearly modulated by the doped elements for their shifting peaks. The obvious shifting of peak V demonstrates most $Ga^{3+}$ cations tend to replace the $Fe^{3+}$ cations inside $FeO_6$ octahedra of *R* blocks, while the prominent shifting of peak VI indicates most $In^{3+}$ cations preferentially replace the $Fe^{3+}$ cations inside the $FeO_5$ bipyramids of *R* blocks as shown in Table I.

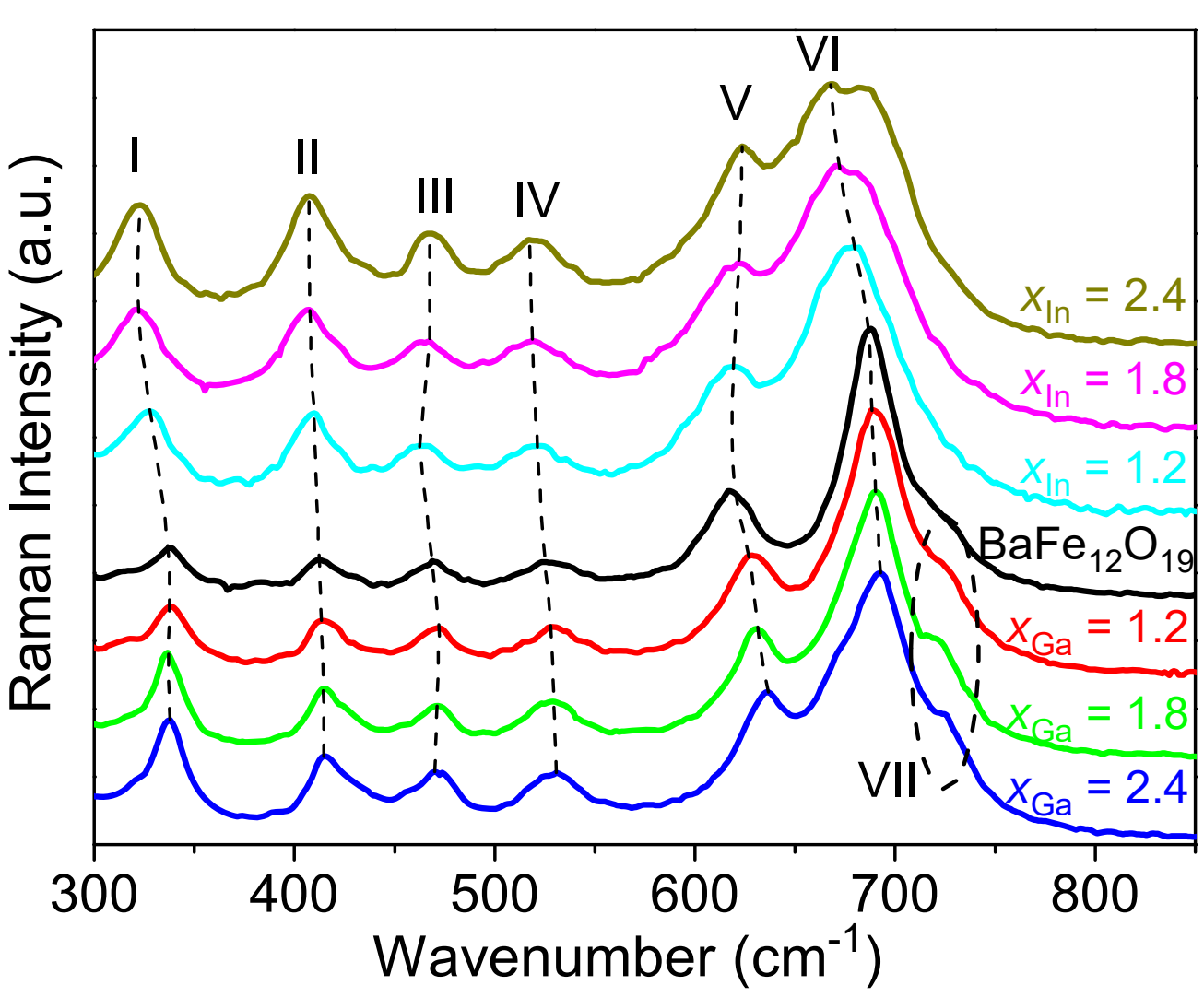

FIG. 3. Raman spectra of $BaFe_{12-x}Me_xO_{19}$ (*Me* = Ga and In; *x* = 0, 1.2, 1.8, and 2.4) ceramics.

Fig. 4 shows the XPS spectra of $BaFe_{12-x}Me_xO_{19}$ ceramics at room temperature.

The binding energy peaks were fitted with the Lorenz-Gaussian function. The Ba $3d_{5/2}$ and $3d_{3/2}$ peaks at around 779.5 eV and 794.9 eV are from the $Ba^{2+}$ cations as shown in Fig. 4(a). Two peaks around 444.2 eV and 451.7 eV correspond to the In $3d_{5/2}$ and In $3d_{3/2}$ of $In^{3+}$ cations as shown in Fig. 4(b). And two peaks around 1117.3 eV and 1144.2 eV are from Ga $2p_{3/2}$ and Ga $2p_{1/2}$ of $Ga^{3+}$ cations as shown in Fig. 4(c).

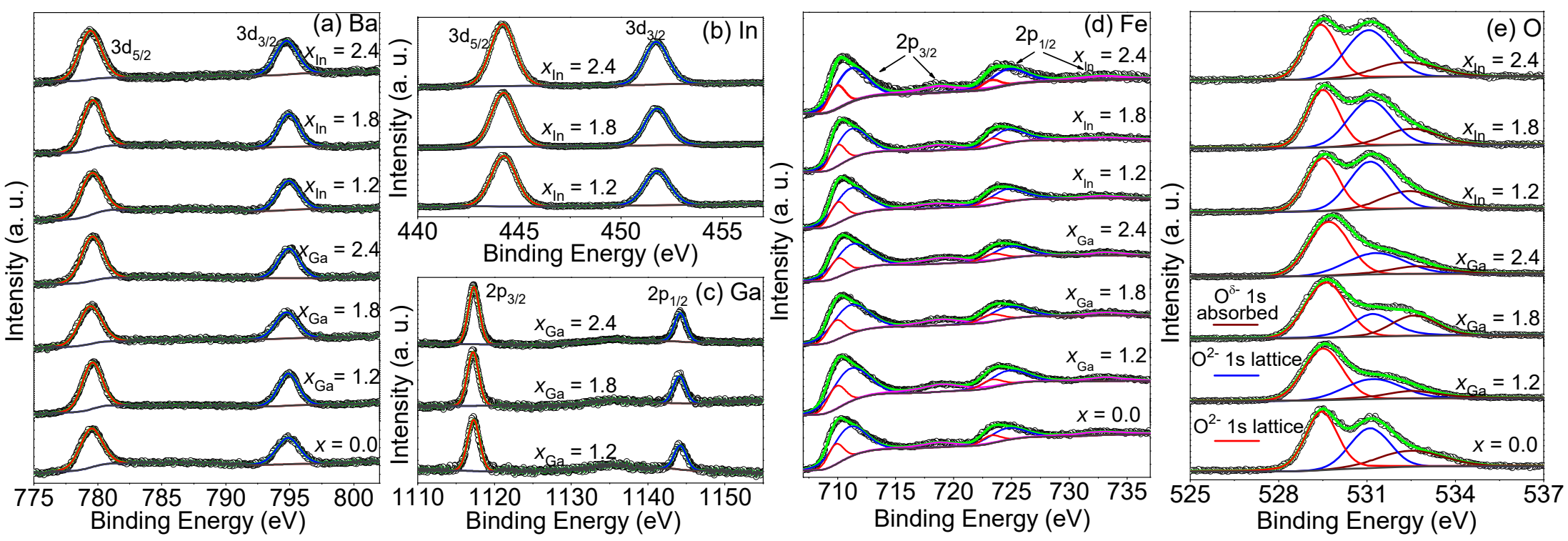

FIG. 4. XPS spectra of (a) Ba 3d, (b) In 3d, (c) Ga 2p, (d) Fe 2p, and (e) O 1s for $BaFe_{12-x}Me_xO_{19}$ (*Me* = Ga and In; *x* = 0, 1.2, 1.8, and 2.4) ceramics.

The XPS patterns of Fe 2p are relatively complex due to its variable chemical states. The main and corresponding satellite peaks of Fe $2p_{3/2}$ and Fe $2p_{1/2}$ are observed for all samples as shown in Fig. 4(d). The main peaks at around 711.0 eV, 724.6 eV, 709.8 eV, and 723.2 eV originate from $Fe^{3+}$ $2p_{3/2}$, $Fe^{3+}$ $2p_{1/2}$, $Fe^{2+}$ $2p_{3/2}$, and $Fe^{2+}$ $2p_{1/2}$ of iron cations, respectively. Correspondingly, the satellite peaks at around 718.6 eV and 732.7 eV are from $Fe^{3+}$ $2p_{3/2}$ and $Fe^{3+}$ $2p_{1/2}$, respectively [29,36,37]. The oxygen loss will result in oxygen vacancies in the ceramics prepared at high temperatures. Then, part $Fe^{3+}$ cations will gain electrons via the reduction reaction to be $Fe^{2+}$ cations in the ceramics [38]. These processes can be expressed as

$$\mathrm{Oo} \rightarrow \frac{1}{2}\mathrm{O}_2 + \mathrm{V\ddot{o}} + 2e' \quad (1)$$

$$\mathrm{Fe}^{3+} + e' \leftrightarrow \mathrm{Fe}^{2+} \quad (2)$$

But part $Fe^{3+}$ cations change to $Fe^{2+}$ through generating holes into crystal lattice by thermal excitation. This process is unrelated to oxygen defects. The fitted $Fe^{2+}/Fe^{3+}$ ratio is about 1/3.3 for these samples after annealing at 1000 °C in an oxygen atmosphere, smaller than those of the reported results [29,36,39]. Three O 1s peaks are observed for all samples as shown in Fig. 4(e). The O 1s peaks at around 529.5 eV and 531.1 eV are attributed to the $O^{2-}$ anions in *R* and *S* blocks, respectively. The discrepant binding energy results from their different chemical environments [40]. The O 1s peak at around 532.5 eV is ascribed to the surface-adsorbed oxygen.

## B. Magnetic property

The thermomagnetic curves of $BaFe_{12-x}Me_xO_{19}$ (*Me* = Ga and In; *x* = 0, 1.2, 1.8, and 2.4) ceramics under zero-field cooling (ZFC) and field cooling (FC) processes were measured from 10 K to 300 K because the magnetic property closely impacts the MD effect. Their obvious differences in Fig. 5 indicate that the doped $Ga^{3+}$ and $In^{3+}$ cations effectively modulate the magnetic property of *M*-type hexaferrite. For the pure $BaFe_{12}O_{19}$ and Ga-doped samples, their ZFC curves exhibit wide peaks at $T_{M1}$ as shown in Figs. 5(a)–5(d). Their magnetization in FC curves monotonically decreases from 10 K to 300 K, exhibiting a typical ferrimagnetic feature without magnetic phase transition [29,41]. The collapse of ferrimagnetic structure is the main factor for this decrease in magnetization.

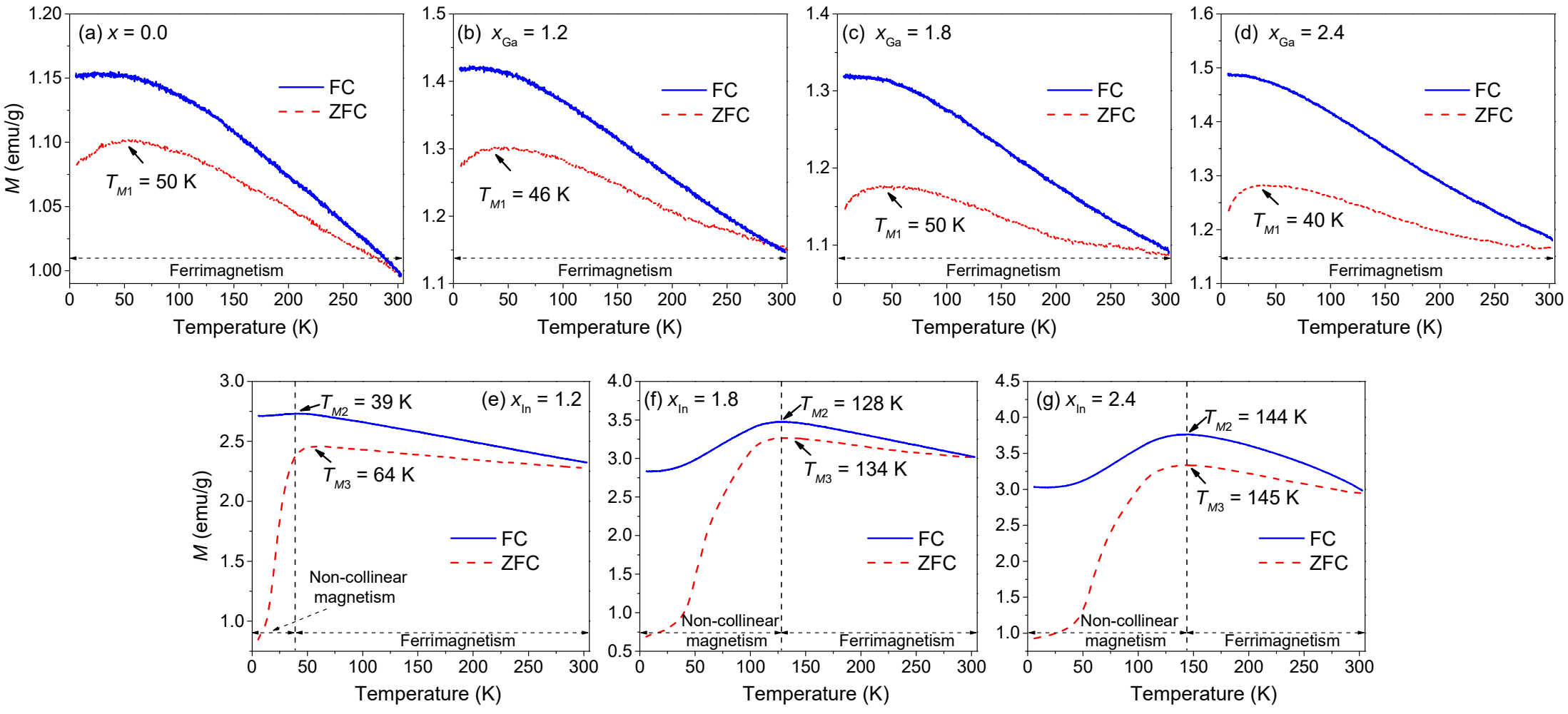

FIG. 5. Thermomagnetic curves of $BaFe_{12-x}Me_xO_{19}$ (*Me* = Ga and In; *x* = 0, 1.2, 1.8, and 2.4) ceramics. The testing field is 100 Oe and the FC field is also 100 Oe.

On the contrary, obvious peaks appear at $T_{M2}$ in the FC curves of In-doped samples as shown in Figs. 5(e)–5(g) and Table I. These peaks reflect the phase transformation from non-collinear magnetism to ferrimagnetism. This transformation is attributed to the spin rotation from the direction away from the *c*-axis [Fig. 1(c)] to the direction along the $c$-axis [Fig. 1(b)], increasing the magnetization from 10 K to $T_{M2}$ [10]. The doped $In^{3+}$ cations in $FeO_5$ bipyramids of $R$ blocks are crucial for non-collinear magnetism [42,43]. The substitution for $Fe^{3+}$ cations inside $FeO_5$ bipyramids will break the original magnetic exchange interaction and then, the antisymmetric exchange becomes important between the magnetic ions on either side of (0 0 1) mirror plane, where the $Ba^{2+}$ cation is located. Besides, the $Fe^{3+}$ cations inside $FeO_5$ bipyramids contribute to the uniaxial magnetocrystalline anisotropy with the $c$-axis as an easy axis. The substitution for $Fe^{3+}$ cations inside $FeO_5$ bipyramids will cause the diversion of easy magnetization axis from the $c$-axis to the $ab$ plane. Consequently, the magnetic structure has a different spin configuration between $L$ and $L$* blocks after the preferential replacement of the $Fe^{3+}$ cations in $FeO_5$ bipyramids by $In^{3+}$ cations as shown in Fig. 1(b). The ionic spins inside a magnetic block are collinear, but the net moments of magnetic blocks ($Spin_L$ and $Spin_{L*}$) are mutually canted and deviated from the $c$-axis, establishing a non-collinear longitudinal conical magnetism as shown in Fig. 1(c). Besides, the increase in $T_{M2}$ indicates the reinforced non-collinear magnetic structure with the In-doping amount [20,21]. From $T_{M2}$ to 300 K, the decrease in magnetization results from the ferrimagnetic destruction by thermal disturbance. There are distinct differences between the ZFC and FC curves for the In-doped samples. The temperature $T_{M3}$ corresponding to maximal magnetization increases with the In-doped amount, and the FC magnetization changes more remarkably than ZFC magnetization below $T_{M3}$. After doping $In^{3+}$ cations, the magnetic interaction weakens, causing the magnetic nanodomains. The frozen magnetic nanodomains lead to a spin-glass state at low temperatures, resulting in the differences between ZFC and FC curves [20,21].

Fig. 6 shows the hysteresis loops of $BaFe_{12-x}Me_xO_{19}$ ceramics at some selected temperatures to comprehensively understand their magnetic properties. The insets show the initial magnetization curves and hysteresis loops at 10 K, whose relative positions reflect the magnetic characteristics. The initial magnetization curves of pure $BaFe_{12}O_{19}$ and Ga-doped samples are located inside their hysteresis loops, coinciding with the normal ferrimagnetic materials. But the initial magnetization curves of In-doped samples reach saturation from outside their hysteresis loops, exhibiting a typical characteristic of non-collinear magnetic structure [44,45].

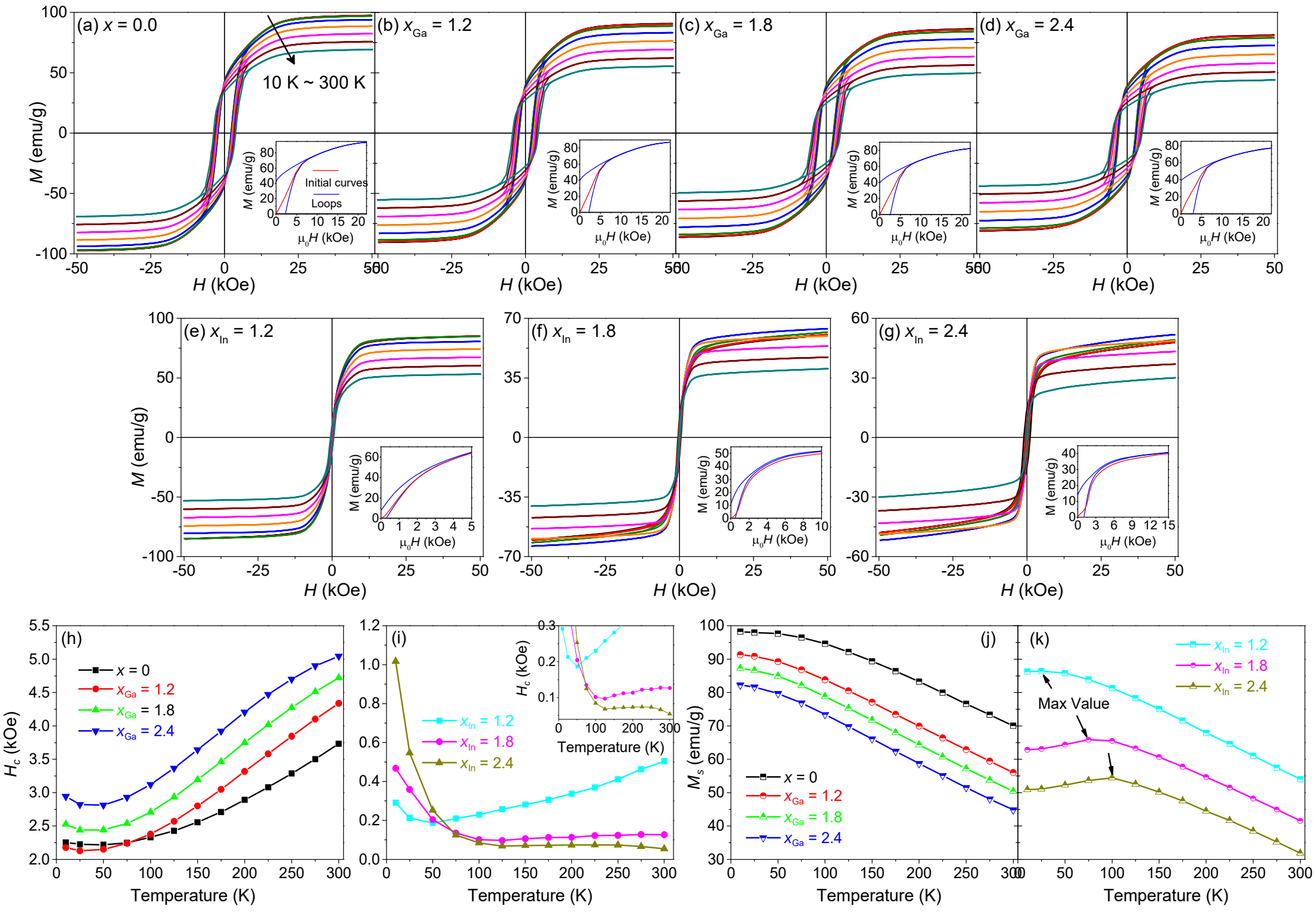

FIG. 6. (a)–(g) Hysteresis loops of $BaFe_{12-x}Me_xO_{19}$ (*Me* = Ga and In; *x* = 0, 1.2, 1.8, and 2.4) ceramics at various temperatures. The insets show the initial magnetization curves and hysteresis loops at 10 K. (h) and (i) Temperature dependence of ceramic coercivity. The inset shows the blown-up patterns. (j) and (k) Temperature dependence of saturation magnetization.

The coercivity ($H_c$) collected from hysteresis loops is shown in Fig. 6(h) for the pure $BaFe_{12}O_{19}$ and Ga-doped samples. $H_c$ firstly decreases and then increases with the temperature, and it increases integrally with the doping amount. $H_c$ is relative to the magnetic domain switching under an external magnetic field, depending on the magnetocrystalline anisotropy and domain wall movement [33]. The initial magnetization ($M$) and magnetic field ($H$) of polycrystalline in strong magnetic fields follow [41]:

$$M = M_s\left[1 - \frac{A}{H} - \left(\frac{B}{H}\right)^2 - \ldots\right] + \chi H \tag{3}$$

where $M_s$ is the saturation magnetization. $A$ and $B$ are determined by the inhomogeneity and magnetocrystalline anisotropy of polycrystalline, respectively, and $\chi$ is the paramagnetic susceptibility at high fields. The magnetization in the range of 43–50 kOe can be plotted against $1/H^2$, i.e., $M = M_s\left[1-(B/H)^2\right]$. The saturation magnetization ($M_s$) and $M_sB^2$ can be obtained by the intercept on the $y$-axis and the slope, respectively. Then, the effective magnetocrystalline anisotropy constant ($K_{eff}$) and magnetocrystalline anisotropy field ($H_a$) can be calculated by

$$K_{eff} = \frac{\sqrt{15}}{2} M_s B \tag{4}$$

$$H_a = \frac{2K_{eff}}{M_s} \tag{5}$$

For the pure $BaFe_{12}O_{19}$ and Ga-doped samples, $H_c$ exhibits a similar trend to $H_a$ with temperature and doping amount because $H_c$ is in proportion to $H_a$ [28], indicating magnetocrystalline anisotropy plays a major role in $H_c$ for these samples.

Fig. 6(i) shows the coercivity of In-doped samples. There are clear temperature turnings at 50 K, 125 K, and 125 K for the In-doped samples with $x_{\mathrm{In}}$ = 1.2, 1.8, and 2.4, respectively. $H_c$ firstly decreases and then increases with temperature. The influences on $H_c$ are quite different above and below the temperature turnings. Below the temperature turnings, the In-doped samples have a non-collinear magnetic structure [Figs. 5(e)–5(g)]. Generally, the non-collinear hexaferrite shows small $H_c$ due to its weak magnetocrystalline anisotropy with the easy magnetization axis away from the $c$-axis [42]. But for the In-doped samples in this work, the coercivity in non-collinear magnetic phase is higher than that in the ferrimagnetic phase, meaning the non-collinear magnetism is not the prominent factor. In fact, the nanomagnetic domains appear after In-doping. More frozen nanomagnetic domains will impede the domain wall movement and lead to an increase in $H_c$ at low temperatures, especially for the hexaferrite with high indium dopant content. Nevertheless, $H_c$ of In-doped samples is still crucially affected by the weak magnetocrystalline anisotropy of non-collinear magnetism, becoming remarkably lower than that of the pure $BaFe_{12}O_{19}$ and the Ga-doped samples with ferrimagnetism at low temperatures. Above the temperature turning, the increase in nanomagnetic domains reduces the magnetocrystalline anisotropy, decreasing $H_c$ after doping $In^{3+}$ cations [20].

$M_s$ decreases with temperature for the pure $BaFe_{12}O_{19}$ and Ga-doped samples as shown in Fig. 6(j), similar to their FC thermomagnetic curves. The weakening magnetic interaction caused by thermal disturbance leads to the $M_s$ reduction. The replacement of partial $Fe^{3+}$ cations in spin-up sites by nonmagnetic $Ga^{3+}$ cations will destroy the original magnetic interaction, resulting in the $M_s$ reduction [22].

The In-doped samples exhibit maximal $M_s$ at certain temperatures as shown in Fig. 6(k), which shares a similar characteristic with their FC thermomagnetic curves due to the magnetic phase transition from non-collinear magnetism to ferrimagnetism. The maximal $M_s$ appears at 25 K, 75 K, and 100 K for the $BaFe_{12-x}In_xO_{19}$ samples with $x_{In}$ = 1.2, 1.8, and 2.4, respectively. The distinction between these temperatures and $T_{M2}$ is attributed to the different strengths of applied magnetic field during the measurements of thermomagnetic curves and hysteresis loops. The replacement of partial $Fe^{3+}$ cations in spin-up sites by nonmagnetic $In^{3+}$ cations also weakens the magnetic interaction.

**C. Dielectric property**

The dielectric property is another crucial aspect to understand the MD effect besides the magnetism. Fig. 7 displays the temperature dependence of the dielectric permittivity of $BaFe_{12-x}Me_xO_{19}$ ($Me$ = Ga and In; $x$ = 0, 1.2, 1.8, and 2.4) ceramics from 10 K to 300 K. The temperature is divided into the low temperature (D-LT) range and close to room temperature (D-CRT) range according to their dielectric features.

The samples exhibit different dielectric characteristics in the D-LT range. The pure $BaFe_{12}O_{19}$ shows a monotonic dielectric decrease from 10 K to about 175 K as shown in Fig. 7(a), attributed to the quantum paraelectric effect [24]. In ferrites, the $Fe^{3+}$ cations are difficult to form hybridized orbitals with their surrounding coordination $O^{2-}$ anions because the $Fe^{3+}$ 3$d$ orbitals are filled with electrons, impeding the deviation of $Fe^{3+}$ cation from the charge center [46]. But this situation is different in the $M$-type hexaferrite. The Pauling repulsion drives $Fe^{3+}$ cations to deviate from their charge center due to the special structure of $FeO_5$ bipyramid, achieving the electric dipoles inside $FeO_5$ bipyramids. *M*-type $BaFe_{12}O_{19}$ hexaferrite is identified as a frustrated dipole system [25]. The electric dipoles inside $FeO_5$ bipyramids distribute triangularly in each *R* or *R** block. Because the *R*/*R** blocks are separated by the *S*/*S** blocks, the distribution of electric dipoles inside $FeO_5$ bipyramids has a 2D triangular crystal feature. The electric dipoles inside $FeO_5$ bipyramids favor antiferroelectric interactions in combination with the 2D triangular crystal feature, inducing dipole frustration. The frustration will lead to an electric dipole glass state inside $FeO_5$ bipyramids. But this glass state is disturbed by the quantum fluctuation, as a result, the quantum paraelectric effect appears [47].

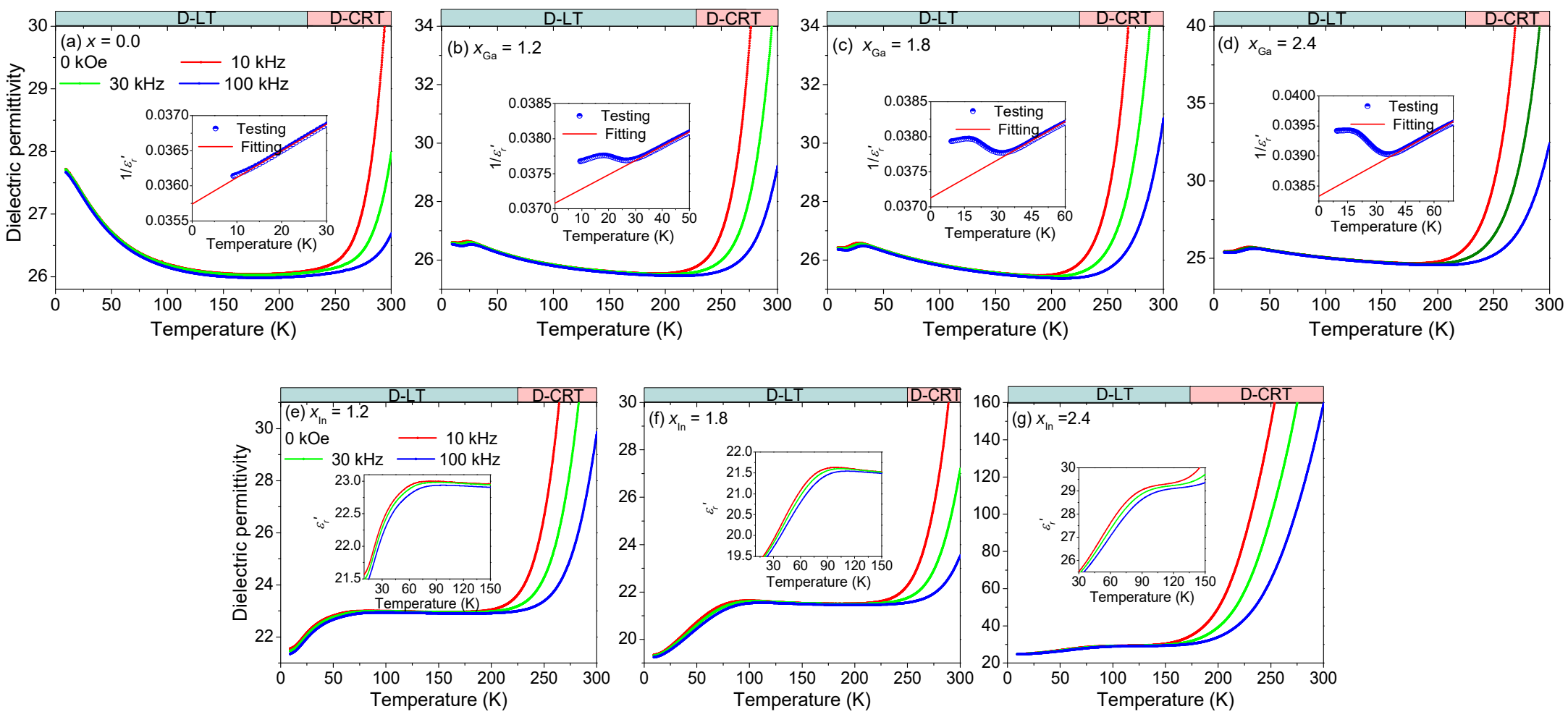

FIG. 7. Temperature dependence of dielectric permittivity of $BaFe_{12-x}Me_xO_{19}$ (*Me* = Ga and In; *x* = 0.0, 1.2, 1.8, and 2.4) ceramics with different frequencies. The insets in (a)–(d) show the temperature dependence of inverse dielectric permittivity and lines fitted by the Curie-Weiss law. The insets in (e)–(g) show blown-up patterns.

Fig. 8 presents the temperature dependence of the dielectric permittivity, tan $\delta$, and $M''$ in the frequency range of 1 Hz–100 kHz from 120 K to 300 K to further investigate the dielectric property. The pure $BaFe_{12}O_{19}$ shows shoulder peaks of tan $\delta$ and $M''$ in the D-LT range as shown in Fig. 8(a), which originates from the electron hopping between adjacent $Fe^{2+}$ and $Fe^{3+}$ cations. The electron hopping will be thermally activated with the increasing temperature to generate defect-dipole ($Fe^{2+}$-$Fe^{3+}$) in the pure $BaFe_{12}O_{19}$ [29]. The electron hopping is random and the defect-dipoles are randomly arranged without electric field. Under the external electric field, the electron hopping will be along the direction of electric field, leading to ordering defect-dipoles, contributing to polarization [38,48]. This polarization has a relaxation characteristic, exhibiting the peaks of tan $\delta$ and $M''$ shift to higher temperatures with the increase in frequency. As shown in Fig. 8(a), the peaks at 1 Hz–1 kHz show a shoulder-like type with little shifts, but are hardly observed at 10–100 kHz, indicating the weak electron hopping in D-LT range for the pure $BaFe_{12}O_{19}$. The shoulder peaks of tan $\delta$ and $M''$ at 10–100 kHz will appear around 180–200 K according to the relaxation characteristic, which is the typical temperature of electron hopping between $Fe^{2+}$ and $Fe^{3+}$ cations in *M*-type hexaferrite [36]. Besides, the polarization from electron hopping will usually cause a steplike increase in dielectric permittivity [29]. But there is no steplike increasing dielectric permittivity in Fig. 7(a) and Fig. 8(a) due to the weak electron hopping in pure $BaFe_{12}O_{19}$.

The electric dipoles inside $FeO_5$ bipyramids in *M*-type $BaFe_{12}O_{19}$ provide a possibility for the dipole glass state [25]. The $Ga^{3+}$ cations tend to replace the $Fe^{3+}$ ions in $FeO_6$ octahedra of *R* block, maintaining the electric dipole and frustration inside $FeO_5$ bipyramids. The Ga-doped samples exhibit a dielectric peak in the range of 20–40 K as shown in Figs. 7(b)–7(d), which is considered from the dipole glass state. The expansion of the *c*-axis will reduce the Coulombic repulsion to promote the ferroelectricity in magnetic compounds [49,50], suppressing the quantum fluctuation to accomplish a transition from quantum paraelectricity to a more ordered state for the electric dipoles inside $FeO_5$ bipyramids. But the electric dipoles inside $FeO_5$ bipyramids for the Ga-doped *M*-type hexaferrite are in a glass state in the low-temperature range rather than ferroelectric or antiferroelectric states because of the frustration. The Curie-Weiss temperatures ($T_{CW}$) fitted by the Curie-Weiss law are −943.5 K, −1825.2 K, −2054.8 K, and −2163.9 K for the pure $BaFe_{12}O_{19}$ and Ga-doped samples as shown in the insets of Figs. 7(a)–7(d). The increase in absolute $|T_{CW}|$ after doping $Ga^{3+}$ indicates the reinforced interaction among electric dipoles inside $FeO_5$ bipyramids. The Ga-doped samples do not show the shoulder peaks of tan $\delta$ and $M''$ from the electron hopping in the D-LT range as shown in Figs. 8(b)–8(d), which is different with the pure $BaFe_{12}O_{19}$. The preferential octahedrally-coordinated locations of $Ga^{3+}$ cations will obviously weaken the electron hopping, because this hopping occurs between the adjacent $Fe^{2+}$ and $Fe^{3+}$ cations in octahedra [51-53], suppressing the shoulder peaks of $\tan\delta$ and $M''$.

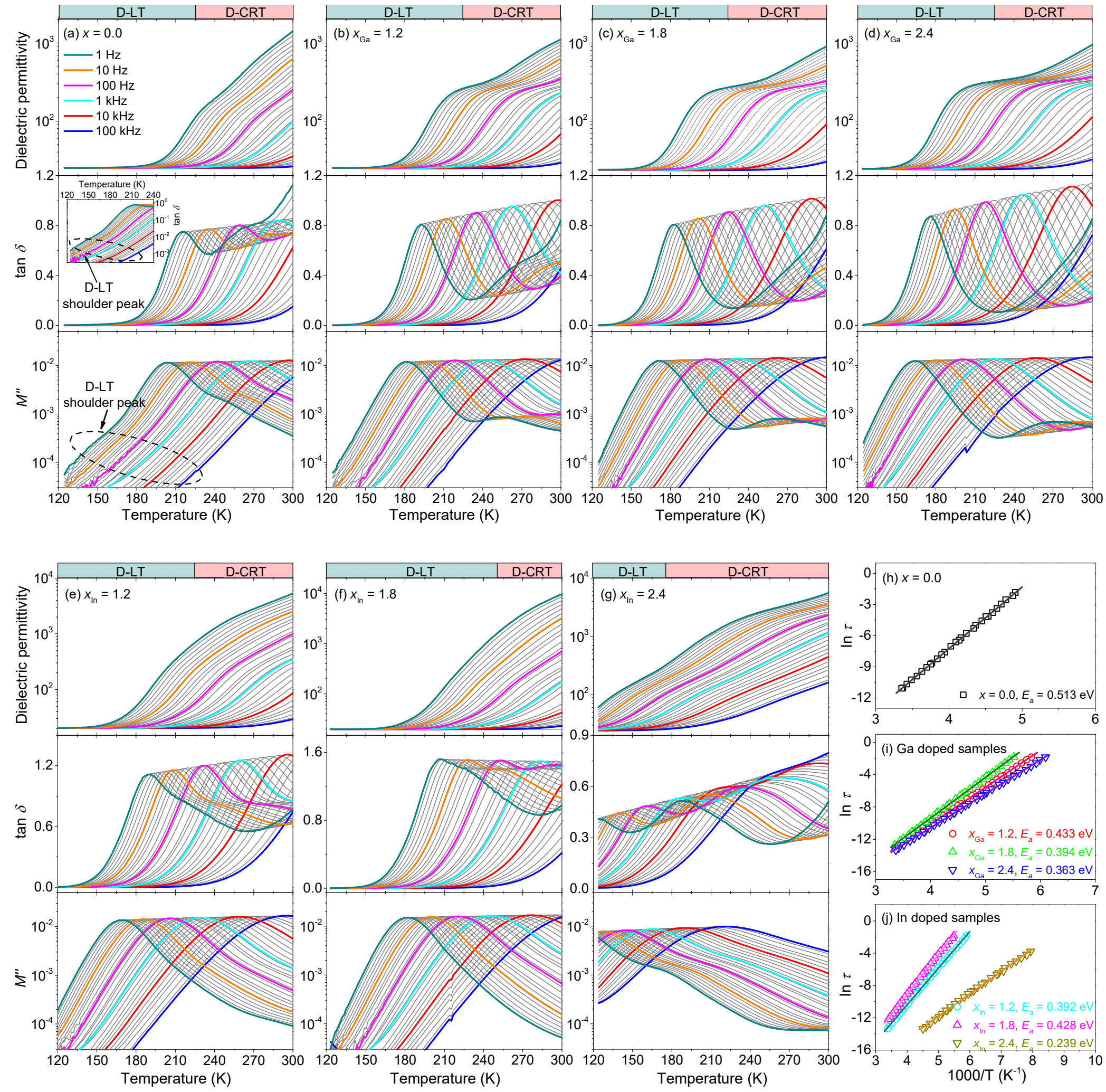

FIG. 8. (a)–(g) Temperature dependence of dielectric permittivity, dielectric loss ($\tan\delta$), and imaginary of the complex electric modulus ($M''$) of $BaFe_{12-x}Me_xO_{19}$ ($Me$ = Ga and In; $x$ = 0.0, 1.2, 1.8, and 2.4) ceramics at different frequencies. (h)–(j) Arrhenius plots from $M''$. The inset in (a) shows the blown-up pattern of dielectric loss in exponential form.

There is a plateau in the dielectric permittivity of In-doped samples as shown in Figs. 7(e)–7(g). The $In^{3+}$ cations with larger radius preferentially replace the $Fe^{3+}$ ions inside the $FeO_5$ bipyramids of $R$ blocks. The off-centered deviation will decrease after doping $In^{3+}$ cations due to the change of interaction between the $In^{3+}$ cations and their surrounding $O^{2-}$ anions of $FeO_5$ bipyramids [32], resulting in the instability or even elimination of the intrinsic electric dipole in bipyramid. Then, the dielectric peak relative to the dipole glass state disappears, and a dielectric plateau from electron hopping appears. The electron hopping between adjacent $Fe^{2+}$ and $Fe^{3+}$ cations in octahedra will not weaken because the $In^{3+}$ cations tend to replace the $Fe^{3+}$ cations inside $FeO_5$ bipyramids. In addition, the extensive unit cell after In-doped will extend the electron hopping distance, causing larger defect-dipoles. Thus, the dielectric contribution from electron hopping is enhanced after doping $In^{3+}$ cations in comparison with the pure $BaFe_{12}O_{19}$, leading to a rapid increase in 20–90 K and then a plateau in 90–200 K of dielectric permittivity.

In the D-CRT range, the obvious dielectric increase and notable frequency dispersion result from the interfacial polarization caused by the Maxwell-Wagner dielectric relaxation. The charge carriers in hexaferrite, including oxygen defects, weakly bound electrons, and ions are activated in this temperature range. These charge carriers will move under an alternating electric field. In ceramics, the grain boundaries with relatively high resistivity are barriers to the movement of charge carriers to concentrate them, generating an interfacial polarization to cause the frequency-dependent dielectric permittivity at low frequencies [54,55].

The Maxwell-Wagner dielectric relaxation clearly shows the peaks of tan $\delta$ and $M''$ in the D-CRT range for all samples as shown in Fig. 8, corresponding to the rapid increase in dielectric permittivity. The complex electric modulus ($M^*$) is less influenced by the increase in conductivity than the complex dielectric permittivity ($\varepsilon^*$) because $M^*$ is the reciprocal of $\varepsilon^*$ as

$$M^* = M' + jM'' = \frac{1}{\varepsilon^*} = \frac{\varepsilon'}{\varepsilon'^2 + \varepsilon''^2} + j\frac{\varepsilon''}{\varepsilon'^2 + \varepsilon''^2} \tag{6}$$

where $\varepsilon'$, $\varepsilon''$, $M'$, and $M''$ are the real part of $\varepsilon^*$, imaginary part of $\varepsilon^*$, real part of $M^*$, and imaginary part of $M^*$, respectively. Therefore, we chose the $M''$ peaks to investigate the dielectric mechanism. The $M''$ peaks move to higher temperatures with the frequency, and the relaxation times ($\tau$) can be described by the Arrhenius law [38]

$$\tau = \tau_0 \exp\left(E_a / k_B T\right) \tag{7}$$

where $\tau$ is $\tau = 1/(2\pi f)$, $f$ is the testing frequency, $\tau_0$ is the factor, $E_a$ is the activation energy of dielectric relaxation, and $k_B$ is the Boltzmann constant. Figs. 8(h)–8(j) shows the linear fitting and the obtained activation energies match the Maxwell-Wagner effect in $M$-type hexaferrite [56].

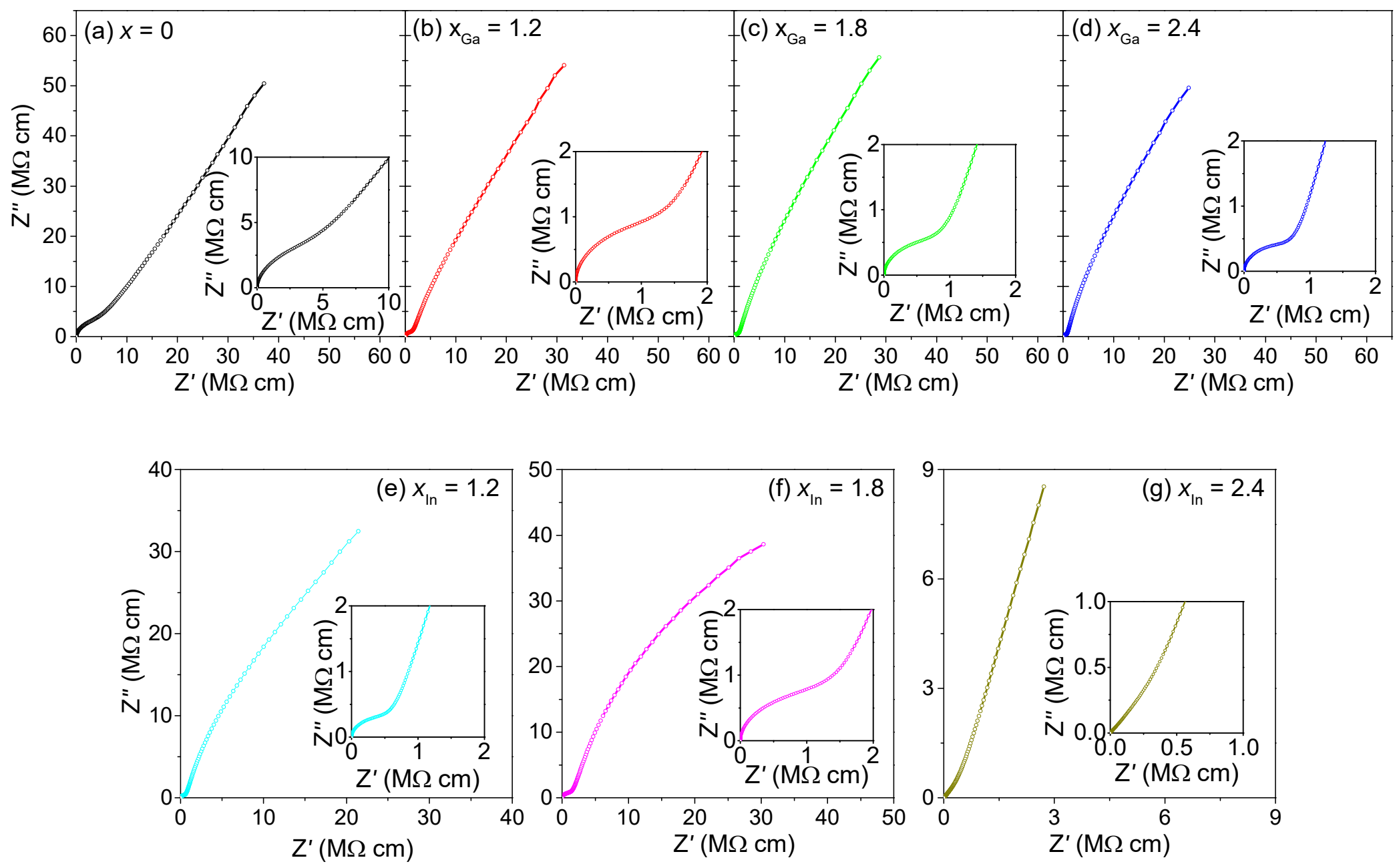

FIG. 9. Impedance complex plots of $BaFe_{12-x}Me_xO_{19}$ ($Me$ = Ga and In; $x$ = 0.0, 1.2, 1.8, and 2.4) ceramics at room temperature. The insets show blown-up patterns at low impedance.

As shown in Fig. 9, there are two arcs with larger and smaller radii at low and high frequencies in the impedance spectra at room temperature, which result from the insulative grain boundary and conductive grain, respectively, supporting the interfacial polarization in the D-CRT range in Fig. 7 [57]. Table I shows the dielectric mechanisms in various temperature ranges for all samples.

## D. Magnetodielectric effect

Fig. 10 shows the magnetic field dependence of the MD coefficient at various temperatures for the $BaFe_{12-x}Me_xO_{19}$ ($Me$ = Ga and In; $x$ = 0, 1.2, 1.8, and 2.4) ceramics and their blown-up patterns at low magnetic fields. The MD coefficient is described as

[58]:

$$\frac{\Delta\varepsilon_r'}{\varepsilon_r'(50\ \text{kOe})} = \frac{\varepsilon_r'(H) - \varepsilon_r'(50\ \text{kOe})}{\varepsilon_r'(50\ \text{kOe})} \tag{8}$$

where $\varepsilon_r'$ ($H$) and $\varepsilon_r'$ (50 kOe) are the dielectric permittivity at variable magnetic fields and maximal magnetic field (50 kOe), respectively. The temperature is divided into the negative MD in low-temperature range (MD-LTN), transition MD in low-temperature range (MD-LTT), positive MD in low-temperature range (MD-LTP), and MD close to room temperature range (MD-CRT) according to their MD features. The samples show various behaviors in the MD-LT range but share similar characteristics in the MD-CRT range.

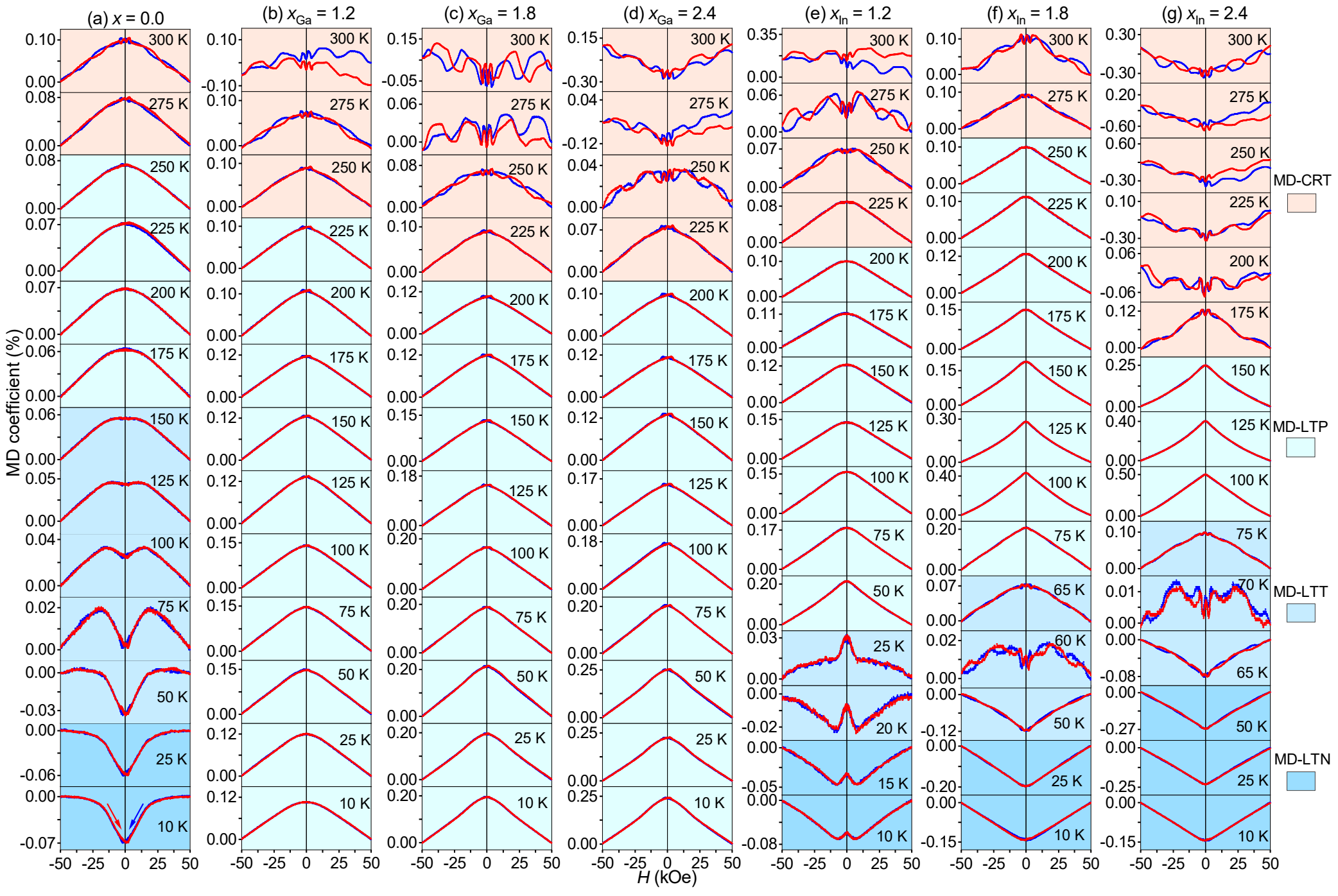

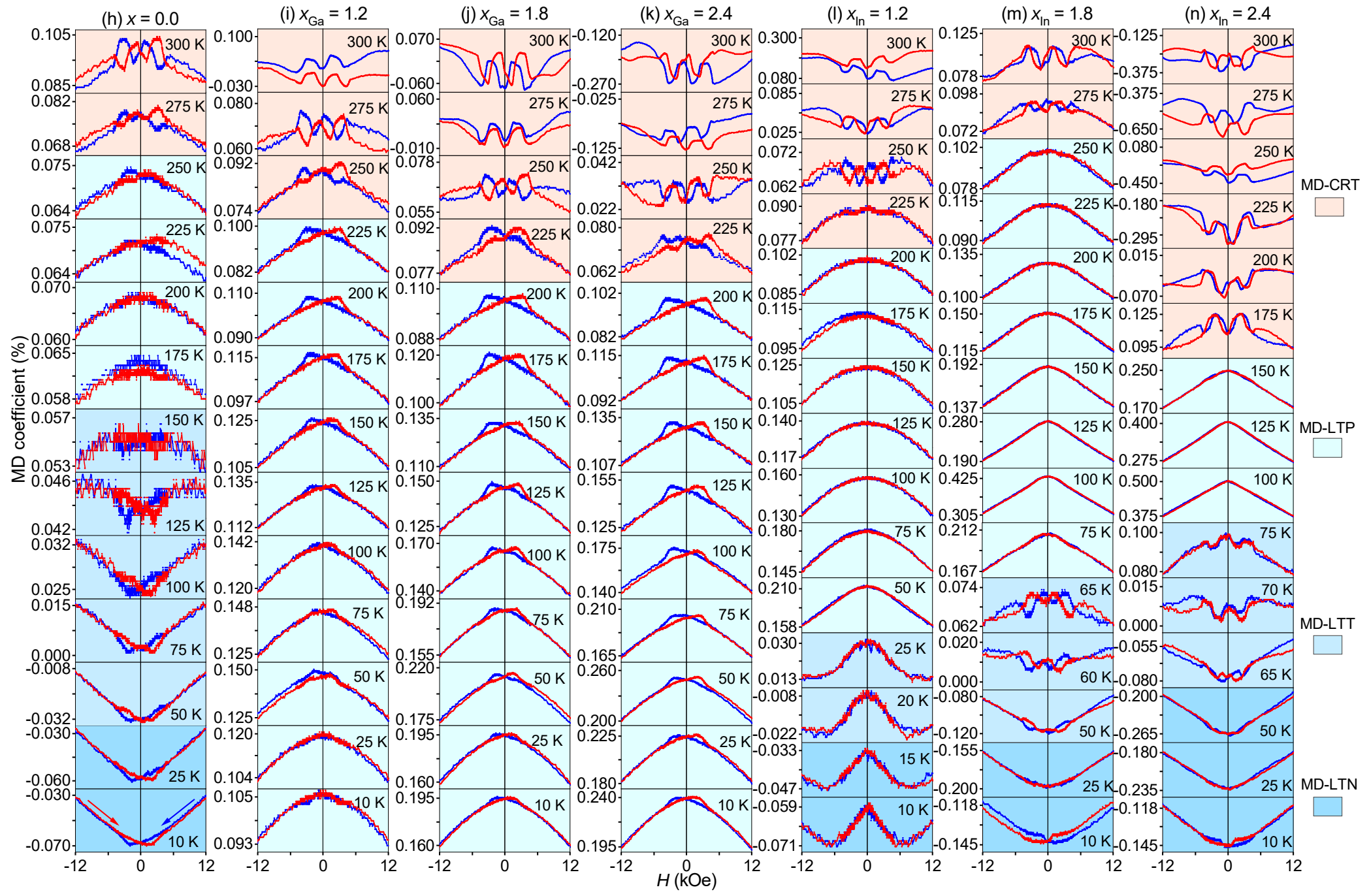

FIG. 10. (a)–(g) Magnetic field dependence of MD coefficient under various temperatures at 100 kHz for the $BaFe_{12-x}Me_xO_{19}$ (*Me* = Ga and In; *x* = 0.0, 1.2, 1.8, and 2.4) ceramics. (h)–(n) their blown-up patterns at low magnetic fields. The blue curves show the MD change from 50 kOe to −50 kOe while the red curves show a reverse change. The different vertical axes show the MD values clearly.

The pure $BaFe_{12}O_{19}$ presents an MD transition from negative to positive in the MD-LT range. The MD coefficient increases obviously at low magnetic fields and remains steady at high magnetic fields in the MD-LTN range. But the MD coefficient decreases monotonically in the MD-LTP range. And the MD coefficients show an MD transition from negative to positive in the MD-LTT range. All Ga-doped samples show a positive MD coefficient in the MD-LTP range. The MD coefficient exhibits a peak around zero field and an inflection point at the middle field. All In-doped samples also present the MD transition from negative to positive in the MD-LT range. In the MD-LTN range, the MD coefficient shows turnings near zero field and then increases with the magnetic field, which is much different from the pure $BaFe_{12}O_{19}$. The MD coefficient decreases monotonically in the MD-LTP range, which is distinct from the Ga-doped samples but similar to the pure $BaFe_{12}O_{19}$. In the MD-LTT range, the MD fluctuations show a transition from negative to positive. All samples exhibit a common feature of irregular MD behaviors in the MD-CRT range. Fig. 11 and Table I summarize the temperature dependence of maximal MD coefficient and the MD coefficients at zero field, showing an obvious modulation by the element doping and temperature.

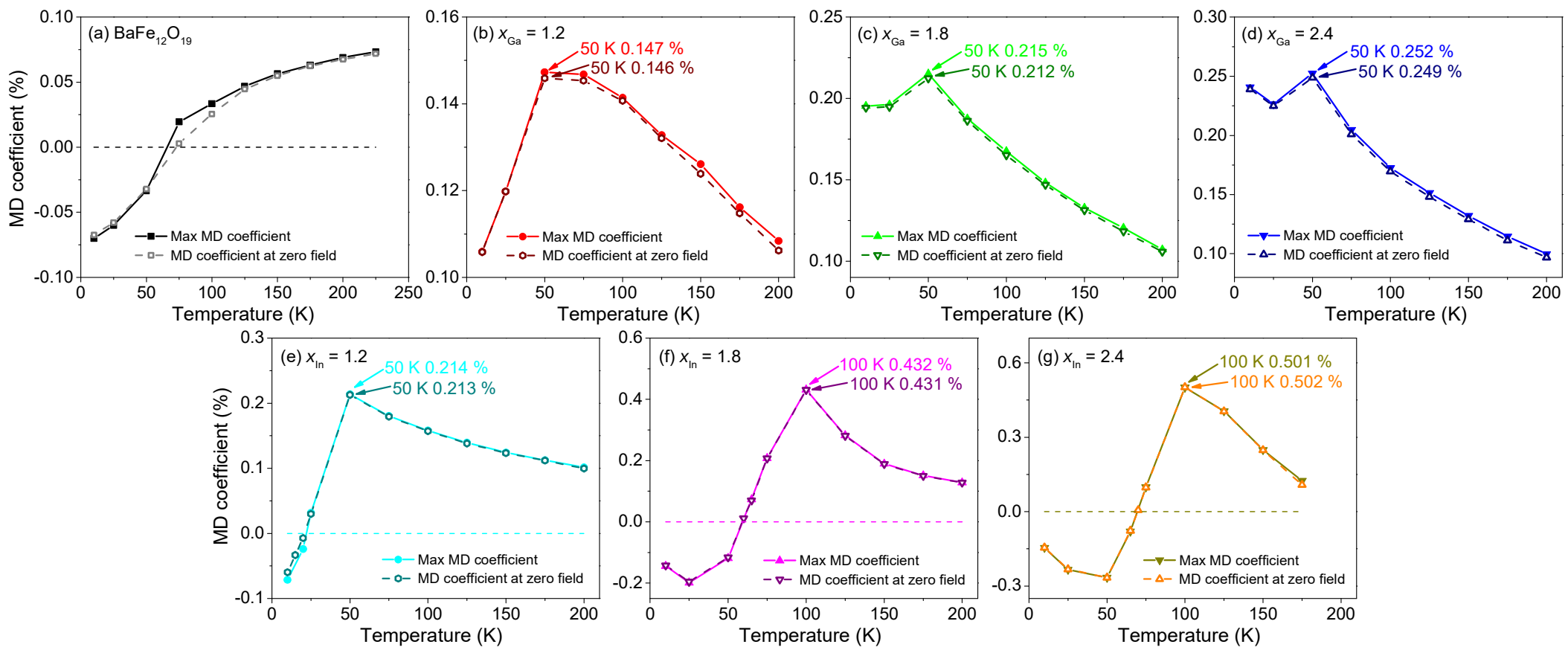

FIG. 11. Temperature dependence of the MD coefficient gathered from the isothermal MD curves. The MD coefficients collected at MD maximum and at zero field share close value.

The MD effect of pure $BaFe_{12}O_{19}$ in the MD-LTN range [Fig. 10(a)] is mainly attributed to the spin-phonon coupling [59]. The lattice vibration, which is characterized by the phonon frequency, can change under a magnetic field through the spin-phonon coupling $\Delta\omega = \lambda\langle \boldsymbol{S}_m \cdot \boldsymbol{S}_n \rangle$, where $\Delta\omega$ is the change in phonon frequencies, $\langle \boldsymbol{S}_m \cdot \boldsymbol{S}_n \rangle$ is a statistical average of adjacent spins, and $\lambda$ is the coupling constant [60]. Then, the spin correlation is built up and hence the spin-phonon coupling becomes important at low temperatures [30]. The dielectric permittivity will change with the phonon frequency on the basis of the Lyddane-Sachs-Teller relationship $\varepsilon_s / \varepsilon_\infty = (\omega_{LO}^2 - \omega^2)/(\omega_{TO}^2 - \omega^2)$, where $\varepsilon_s$ and $\varepsilon_\infty$ are the static permittivity and optical permittivity, respectively. $\omega_{LO}$ and $\omega_{TO}$ are the longitudinal and transverse optical phonon frequencies, respectively. Then, the MD effect is achieved [23,61]. In this scheme, the dielectric difference and the square of magnetization meet $\Delta\varepsilon \sim \gamma M^2$, where $\gamma$ is the coupling coefficient [8,62-64]. Thus, the linear relationship is maintained between $\Delta\varepsilon$ and $M^2$ at the middle magnetic fields as shown in Fig. 12, confirming the MD effect originates from the spin-phonon coupling. The magnetic domain wall movement at low magnetic fields and the paramagnetic magnetization at high magnetic fields play a main part in the magnetization, resulting in a deviation from the linearity between $\Delta\varepsilon$ and $M^2$.

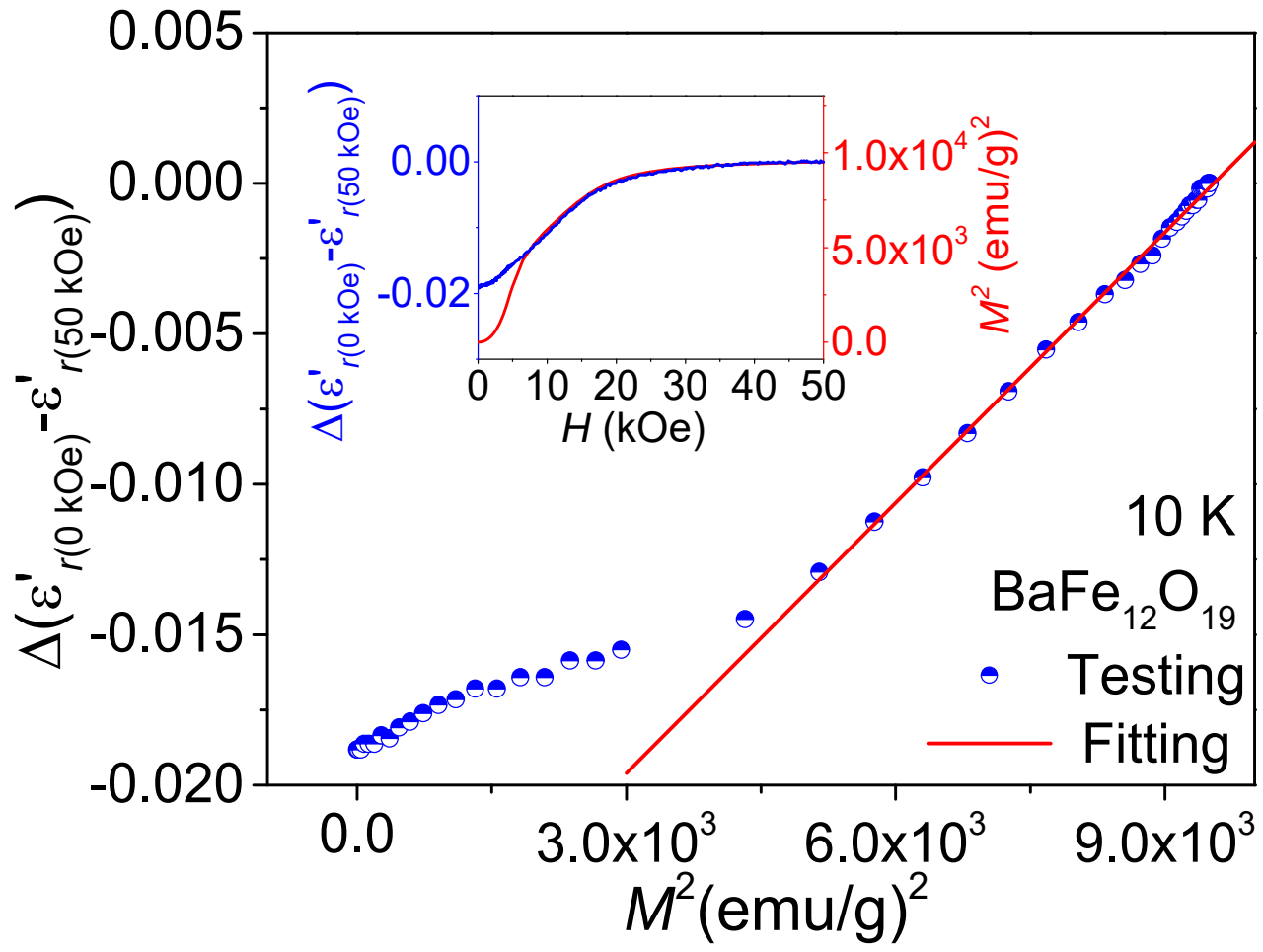

FIG. 12. Dielectric difference as a function of the square of magnetization for the pure $BaFe_{12}O_{19}$ at 10 K. The inset shows the magnetic field-dependent dielectric difference and square of magnetization, where two curves overlap together in a large magnetic field range.

With the increase in temperature, the MD coefficient changes to positive values in the MD-LTP range for the pure $BaFe_{12}O_{19}$ [Fig. 10(a)], which derives from the field-dependent electron hopping in this temperature range [Fig. 8(a)]. The electron hopping will be disturbed by the magnetic field through the Lorentz force, causing a decrease in dielectric permittivity. The gradual change in MD curves in the MD-LTT range indicates this range is the transitional stage. With the increase in temperature, the electron hopping will enhance, leading to the increase in MD coefficients as shown in Fig. 11(a).

The coexistence of ferrimagnetism and electric dipoles inside $FeO_5$ bipyramids provides a foundation for the MD effect in Ga-doped samples. Both the electron spins and electric dipoles are associated with the $Fe^{3+}$ cations in Ga-doped samples, connecting the magnetic and electric properties. The electron spin will turn under the external magnetic field, changing the electron cloud distribution inside $FeO_5$ bipyramid due to the spin-orbit coupling [42]. The electron cloud redistribution modulates the deviation of $Fe^{3+}$ cation from the charge center of $FeO_5$ bipyramid, i.e., the electric dipole inside $FeO_5$ bipyramid. More importantly, the expansion along the *c*-axis after

Ga-doping suppresses the quantum fluctuation and enhances the dipole-dipole interaction in $FeO_5$ bipyramids [65], promoting the coupling of magnetic field and these electric dipoles. In addition, the inflection points in isothermal field-dependent MD coefficient [Figs. 10(b)–10(d)] and the magnetic hysteresis loops [Figs. 6(b)–6(d)] well coincide with each other around 20 kOe. A typical match is exhibited in Fig. 13 for $BaFe_{9.6}Ga_{2.4}O_{19}$ at 10 K. Then, it is credible that the electric dipole inside $FeO_5$ bipyramid is magnetic field-dependent, leading to an MD effect in the MD-LTP range. The maximal MD coefficient around 50 K in Figs. 11(b)–11(d) is relative to the transition from paraelectric state to glass state of the electric dipoles inside $FeO_5$ bipyramids as shown in Figs. 7(b)–7(d).

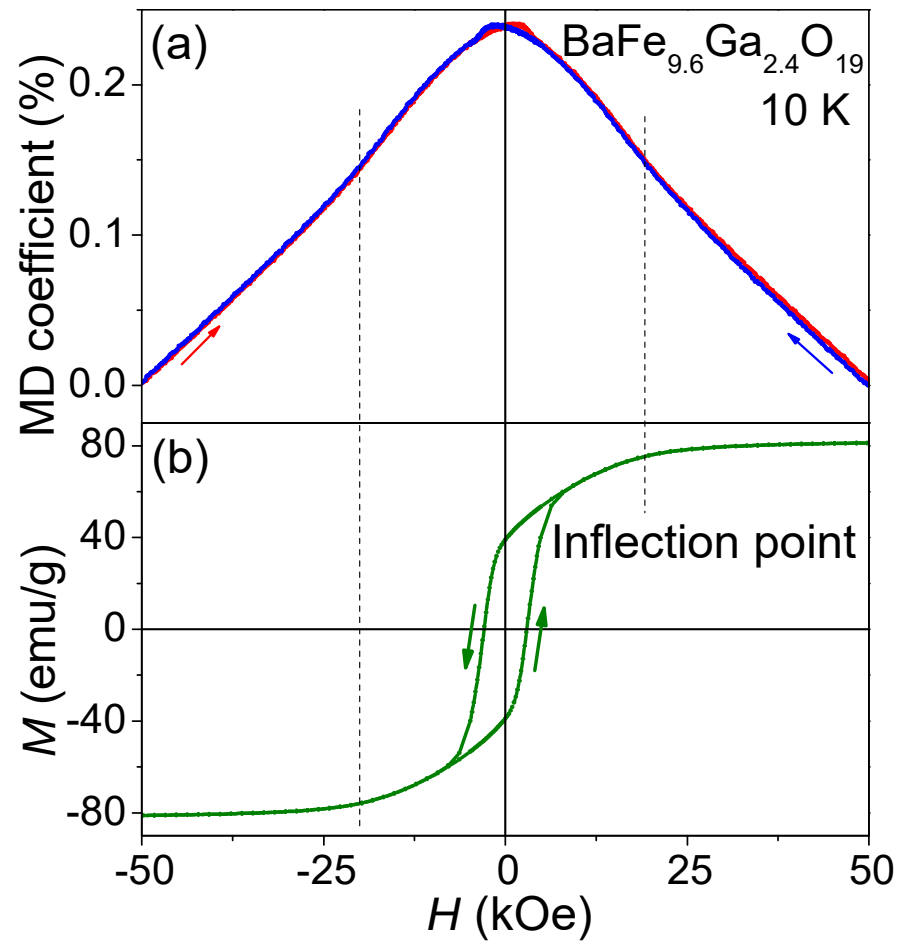

FIG. 13. Magnetic field dependence of (a) MD coefficient and (b) magnetization for the $BaFe_{9.6}Ga_{2.4}O_{19}$ sample at 10 K.

The spin ordering-mediated ME and MD effects are considered from the non-collinear longitudinal conical spin order for the In-doped samples in MD-LTN range. The conical axis of spin order will deviate from the $c$-axis under an external magnetic field, generating electric polarization as shown in Fig. 1(d) [10,20]. This phenomenon can be described by the inverse Dzyaloshinskii-Moriya interaction

$$P \propto \sum \boldsymbol{e}_{ij} \times (\boldsymbol{S}_i \times \boldsymbol{S}_j) \qquad (9)$$

where $\boldsymbol{S}_i$ and $\boldsymbol{S}_j$ are the spins at adjacent sites $i$ and $j$, respectively, and $\boldsymbol{e}_{ij}$ denotes the unit vector connecting them. With the increase in applied magnetic field, the polarization will change with the evolution of spin order as shown in Fig. 1(e). Then, the spin ordering-mediated ME effect is achieved eventually [10,19]. Meanwhile, the polarization from the ME coupling impacts the dielectric permittivity, realizing the spin ordering-mediated MD effect [66]. The In-doped *M*-type hexaferrite represents the spin ordering-mediated ME effect and there are notable mutations in polarization around the low magnetic fields at 10 K [20]. Then, the MD turnings around low magnetic field are related to the ME effect [Figs. 10(l)–10(n)]. Therefore, the MD effect in MD-LTN range is derived from the evolution of spin order under an external magnetic field.

The MD effect in MD-LTP range for the In-doped samples is ascribed to the field-dependent electron hopping [7,29,36]. Fig. 14 shows the temperature dependence of the MD coefficient and dielectric permittivity at different frequencies for the $BaFe_{12-x}In_xO_{19}$ samples. These samples show a typical dielectric relaxation with the rapid dielectric increase and the frequency dependence of dielectric permittivity due to the electron hopping in 20–100 K. The MD transition from negative to positive occurs in this temperature range and the MD peak also shifts to higher temperature. This similar aspect connects the MD effect and electron hopping together. The electron hopping will be disturbed by the magnetic field through the Lorentz force as mentioned in the discussion on pure $BaFe_{12}O_{19}$ in ML-LTP range. In addition, the non-collinear spins tend to align in parallel under an applied magnetic field [Fig. 1(c)], which hinders the electron hopping to modulate the dielectric response. These two factors will decrease the dielectric permittivity of the In-doped samples below $T_{M2}$, while the non-collinear spin order disappears and only the former one makes an influence above $T_{M2}$. With the enhancement of electron hopping in the warming process, the positive MD effect from the field-dependent electron hopping achieves and gradually covers the negative MD effect from field-dependent non-collinear spin ordering in the MD-LTT range. The increase in $T_{M2}$ with the indium amount [Figs. 5(e)–5(g)] indicates the reinforced non-collinear spin ordering, which leads to a wider MD-LTN range and higher change temperature of the MD transition from negative to positive.

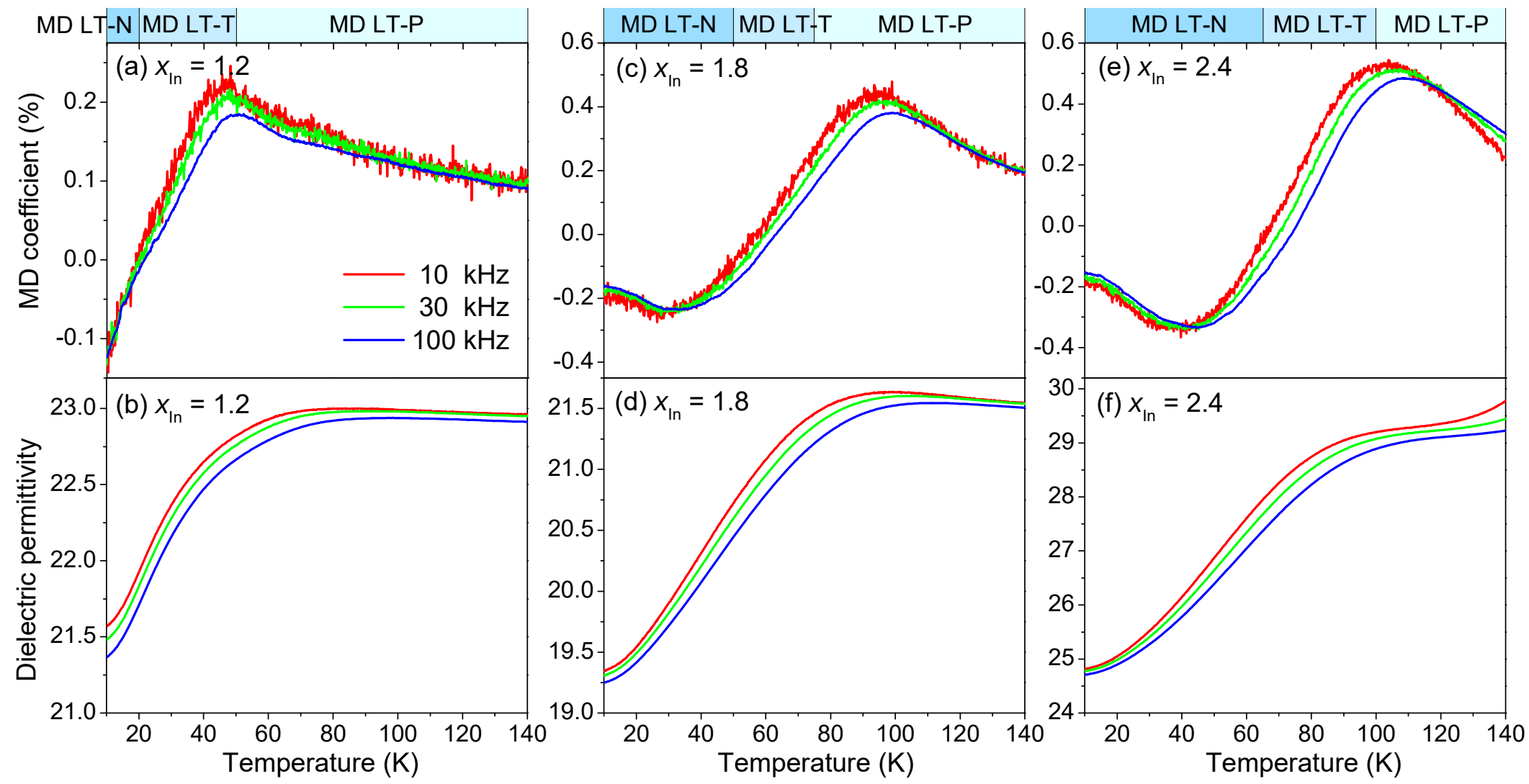

FIG. 14. Temperature dependence of MD coefficient and dielectric permittivity at different frequencies for the In-doped $BaFe_{12-x}In_xO_{19}$ samples with (a, b) $x$ = 1.2, (c, d) $x$ = 1.8 and (e, f) $x$ = 2.4. The MD coefficient is calculated by the dielectric permittivity at 0 kOe and 50 kOe.

In the MD-CRT range, the MD effect mainly originates from the combination of magnetoresistance and Maxwell-Wagner effects due to the rapid dielectric increase from interfacial polarization (Fig. 7). The Maxwell-Wagner effect is a universal dielectric response in polycrystalline materials at low frequencies. The dielectric response to an electric field contains at least one capacitive term and one resistive term. The grains have relatively low resistivity due to the asymmetric electron hopping under an ac electric field. The grain boundaries with relatively high resistivity can impede electron transfer, resulting in charge accumulation at the grain boundaries between different grain regions. This extrinsic source of dielectric relaxation is known as the Maxwell-Wagner relaxation [67]. The impedance of this equivalent circuit consists of two effective leaky capacitors in series

$$Z = \frac{1}{R_G^{-1} + i\omega C_G} + \frac{1}{R_B^{-1} + i\omega C_B} \tag{10}$$

where the subindexes $G$ and $B$ refer to the grain and grain boundary, respectively, and $R$ and $C$ are the resistance and capacitance, respectively. Then, the real and imaginary parts of the effective permittivity are [9]

$$\varepsilon'(\omega) = \frac{1}{C_0(R_G + R_B)} \frac{\tau_G + \tau_B - \tau + \omega^2 \tau_G \tau_B \tau}{1 + \omega^2 \tau^2} \quad (11)$$

$$\varepsilon''(\omega) = \frac{1}{\omega C_0(R_G + R_B)} \frac{1 - \omega^2 \tau_G \tau_B + \omega^2 \tau(\tau_G + \tau_B)}{1 + \omega^2 \tau^2} \quad (12)$$

where $\tau_G = C_G R_G$, $\tau_B = C_B R_B$, $\tau = (\tau_G R_B + \tau_B R_G)/(R_G + R_B)$. Thus, the magnetic field can modulate the dielectric permittivity to produce an MD effect through magnetoresistance. But the magnetoresistance is weak in the *M*-type hexaferrite [15]. Therefore, the MD coefficients are small and unstable with the magnetic field in MD-CRT range, which also suggests that the MD effects at low temperatures are intrinsic behaviors. Table I summarizes the various MD mechanisms for all samples after Ga and In-doping in different temperature ranges.

## IV. Conclusions

This work prepared $BaFe_{12-x}Me_xO_{19}$ (*Me* = Ga and In; $x$ = 0.0, 1.2, 1.8, and 2.4) ceramics by the solid-state reaction. The oxygen defects in ceramics will appear during the high-temperature process. The modulations of crystal, magnetic, and dielectric properties are achieved by doping $Ga^{3+}$ and $In^{3+}$ cations, which also result in MD effects with various mechanisms. The $In^{3+}$ dopants induce the crystal lattice expansion in both *a*-axis and *c*-axis while the $Ga^{3+}$ dopants cause shrinkage in the *a*-axis and expansion in the *c*-axis. The obvious shifting of Raman peaks demonstrates that the $Ga^{3+}$ and $In^{3+}$ cations preferentially substitute the $Fe^{3+}$ ions in $FeO_6$ octahedra and $FeO_5$ bipyramids of *R* blocks, respectively. As a result, the pure $BaFe_{12}O_{19}$ and Ga-doped samples show the usual ferrimagnetism, but the In-doped samples exhibit a non-collinear magnetism at low temperatures, in which their initial magnetization curves are located outside hysteresis loops. The apparent dielectric increase and notable frequency dispersion originate from the Maxwell-Wagner effect for all samples in the D-CRT range. In the D-LT range, the dielectric property shows various features and mechanisms. The monotonic dielectric decrease in 10–175 K is ascribed to the quantum paraelectric effect, and shoulder peaks of tan $\delta$ and $M''$ in 140–200 K result from the electron hopping for the pure $BaFe_{12}O_{19}$ sample. The dielectric peaks of the Ga-doped samples are from the dipole glass state. The dielectric increase and plateau of the In-doped samples are ascribed to the electron hopping. As for the MD effect, the combination of magnetoresistance and Maxwell-Wagner effects leads to a modulation of dielectric permittivity by magnetic field for all samples in the MD-CRT range. In the MD-LT range, the MD effect presents different characteristics and origins. The negative MD effect at extremely low temperatures and the positive MD effect after warming for the pure $BaFe_{12}O_{19}$ are mediated by the spin-phonon coupling and field-dependent electron hopping, respectively. The negative MD effect at extremely low temperatures and the positive MD effect after warming for In-doped samples are controlled by non-collinear spin ordering and electron hopping under a magnetic field, respectively. But this is impossible for the Ga-doped samples because of their ferrimagnetism. The field-dependent electric dipoles inside $FeO_5$ bipyramids are responsible for their positive MD effect in Ga-doped samples. Our research provides helpful results to associate the magnetic, dielectric, and MD properties with the substitution sites of doped ions in *M*-type hexaferrite.

## Acknowledgments

This work was supported by the National Natural Science Foundation of China (Grant No. 51672168), the Fundamental Research Funds for the Central Universities (Grant Nos. 2016TS035, 2017TS010), and the College Students' Innovative Entrepreneurial Training Plan Program (Grant No. S202010718024).